    \documentclass[aps,prb,groupedaddress,twocolumn,amsmath,floatfix]{revtex4}
   \usepackage{graphicx}
\newcommand\beq{\begin{equation}}
\newcommand\eeq{\end{equation}}
\newcommand\beqa{\begin{eqnarray}}
\newcommand\eeqa{\end{eqnarray}}
\newcommand\nn{\nonumber}

\newcommand{\etap}{y}
\newcommand{\yy}{w}

\begin{document}

\title{Equation of state of non-additive $d$-dimensional hard-sphere mixtures}
\author{A. Santos}
\email[]{andres@unex.es}
\homepage[]{http://www.unex.es/fisteor/andres/}
\author{M. L\'opez de Haro}
\email[]{malopez@servidor.unam.mx}
\homepage[]{http://miquiztli.cie.unam.mx/xml/tc/ft/mlh/}
\thanks{On sabbatical leave from Centro de Investigaci\'{o}n en Energ\'{\i}a, U.N.A.M., Temixco, Morelos 62580
(M\'{e}xico)}
\author{S. B. Yuste}\email[]{santos@unex.es}
\homepage[]{http://www.unex.es/fisteor/santos/sby}
\affiliation{Departamento de F\'{\i}sica, Universidad de
Extremadura, E-06071 Badajoz, Spain}
%\thanks{}


\date{\today}

\begin{abstract}
 An equation of state for a multicomponent mixture of
non-additive hard spheres in $d$ dimensions is proposed. It yields a
rather simple density dependence and constitutes a natural extension
of the equation of state for {\em additive} hard spheres proposed by
us [A. Santos, S. B. Yuste, and M. L\'opez de Haro, Mol.\ Phys.\
{\bf 96}, 1 (1999)]. The proposal relies on the known exact second
and third virial coefficients and requires as input the
compressibility factor of the one-component system. A comparison is
carried out both to another recent theoretical proposal based on a
similar philosophy and to the available exact results and simulation
data in $d=1$, $2$, and $3$. {Good general agreement with the
reported values of the virial coefficients and of the
compressibility factor of binary mixtures is observed, especially
for high asymmetries and/or positive nonadditivities.}
\end{abstract}

%\pacs{xxx}
%\keywords{}

\maketitle

\section{Introduction and a brief review of the literature\label{sec1}}
The structure of a dense fluid is known to be largely determined
by the repulsive intermolecular forces, so it is not surprising that
hard-core potentials have been extensively employed to model simple fluids
and fluid mixtures. A noteworthy aspect of these models is the fact that
in some instances both exact and approximate analytical results may be
derived for the structural and thermodynamic properties, which in turn
serve as a starting point for the treatment of more sophisticated or
complex models.

Certainly a vast majority of the published work on hard-core
(rods, disks, spheres, and hyperspheres) fluid mixtures pertains
to binary systems and to the so-called additive hard-core
interaction, namely the one in which the distance of closest
approach (denoted by $\sigma_{ij}$) between the centers of two
interacting particles, one of species $i$ and the other of
species $j$, is the arithmetic mean of the diameters of both
particles $\sigma_i$ and $\sigma_j$, respectively. Apart from the
initial impetus that took place in the 60s, recently interest in
this kind of systems (in particular mixtures of hard spheres) has
experienced an increasing growth in connection with entropy
driven phase transitions and the demixing problem. On the other
hand, non-additive hard-core mixtures, where the distance of
closest approach between particles of different species is no
longer the arithmetic mean referred to above, have received less
attention, in spite of their in principle more versatility to
deal with interesting aspects occurring in real systems (such as
liquid-vapor equilibrium or fluid-fluid phase separation) and of
their potential use as reference systems in perturbation
calculations on the thermodynamic and structural properties of,
say, Lennard-Jones mixtures. Nevertheless, the study of
non-additive systems goes back fifty years\cite{PL54,AO54,K55}
and is still a rapidly developing and challenging problem.

As  mentioned in the paper by Ballone \textit{et al.},\cite{BPGG86}
where the relevant references may be found, experimental work on
alloys, aqueous electrolyte solutions, and molten salts suggests
that hetero-coordination and homo-coordination may be interpreted in
terms of excluded volume effects due to non-additivity of the
repulsive part of the intermolecular potential. In particular,
positive non-additivity leads naturally to demixing in hard-sphere
mixtures, so that some of the experimental findings of phase
separation in the above mentioned (real) systems may be accounted
for by using a model of a binary mixture of (positive) non-additive
hard spheres. On the other hand, negative non-additivity seems to
account well for chemical short-range order in amorphous and liquid
binary mixtures with preferred hetero-coordination.\cite{GPE89}

On the theoretical side, the first exact result on the equation of
state (EOS) for a non-additive mixture is that of a binary mixture
of hard rods ($d=1$) restricted to nearest-neighbor interactions.
Although it is usually atributed to Lebowitz and
Zomick,\cite{LZ71} it was already implicit in earlier work by
Prigogine and Lafleur\cite{PL54} and by Kikuchi,\cite{K55} and
even Lebowitz and Zomick point out that the thermodynamic
functions of this system appear in the thesis presented in 1966
by C. C. Carter ({cf.} Ref.\ 9 in \onlinecite{LZ71}). Very
recently, Corti and Bowles have rederived this result in an
appendix of a paper,\cite{CB99} where they also provide exact
geometrical relationships for non-additive mixtures (see also an
alternative rederivation in Prof.\ Penrose's
webpage\cite{Penrose}).
 It is also worth mentioning that in the paper by
Kikuchi,\cite{K55} a proof is given that no phase transition may
occur in a one-dimensional binary mixture irrespective of the form
of the interaction potential, provided it is unbounded. {The
opposite limit of high spatial dimension has been considered by
Carmesin \textit{et al.},\cite{CFP90} who showed that at
sufficiently high density and with positive nonadditivity, a binary
mixture of non-additive hard hyperspheres decomposes into two
coexisting phases.}

A very popular model of a non-additive binary mixture with positive
non-additivity was introduced by Widom and Rowlinson in
1970.\cite{WR70} This model is equivalent to a one-component
penetrable sphere model. In the symmetric version of the model,
referred to as the Widom--Rowlinson (WR) model, one has
$\sigma_1=\sigma_2=0$ and $\sigma_{12}>0$. With this simplification
Widom and Rowlinson derived  exactly  the EOS in the one-dimensional
case, where it predicts no phase transition. For $d=3$ the model was
solved in the mean-field approximation. In the same paper, but for
the case of high asymmetry (i.e., when $\sigma_1\neq 0$,
$\sigma_2=0$, and $d=3$), Widom and Rowlinson also determined an
approximate condition for the spinodal curve. It is interesting to
point out that this case of high asymmetry corresponds with the
Asakura--Oosawa model,\cite{AO54} often used to discuss polymer
colloid mixtures and where the notion of a depletion potential was
introduced. This model and refinements of the same have received a
lot of attention (including fairly recent work) in connection with
the demixing problem and the question of effective
potentials.\cite{V76}

The impact of the WR model cannot be overemphasized as it has
motivated a great amount of later work. A rigorous proof that a
phase transition may exist in the WR model in $d=2$ was provided by
Ruelle,\cite{R71} who also indicated that a similar procedure may be
followed to prove the existence of a phase transition in the WR
model in $d=3$ and higher dimensions. Frisch and Carlier\cite{FC72}
performed molecular dynamics simulation for a hard-square mixture in
the WR limit and concluded that it presented a first order phase
transition. Melnyck \textit{et al.}\cite{MRS72} obtained the first
ten virial coefficients of the WR hard-sphere mixture in the
Percus--Yevick (PY) approximation (the first five of which are
exact), while Straley \textit{et al.}\cite{SCLW72} computed the
virial coefficients of the WR model for oriented hard squares and
hard cubes. Widom and Stillinger\cite{WS73} generalized the scaled
particle theory (SPT) for a pure fluid\cite{SPT} to the case of the
WR model in an arbitrary dimensionality and Guerrero \textit{et
al.}\cite{GRS74} exploited the equivalence of the penetrable sphere
model and the WR model to obtain the direct and total correlation
functions for the model where the Mayer function is a Gaussian and
for the hard-sphere interaction in the mean field, PY, and
hypernetted chain approximations. In the cases of $d=1$ and $d=3$,
the WR model was solved in the PY approximation by Ahn and
Lebowitz,\cite{AL74}
 while the SPT was considered by Bergmann.\cite{B76a}
The latter theory for the WR model in two dimensions was addressed
in an appendix of the paper by Tenne and Bergmann,\cite{TB79} in
which they examined the SPT for a non-additive hard-disk binary
mixture. Transport properties for the WR hard-sphere binary mixture
were computed by Karkheck and Stell.\cite{KS79}
 Later, Borgelt \textit{et al.}\cite{BHS90} and Luo
\textit{et al.}\cite{LHK90}  performed simulations on the
hard-sphere WR mixture and found better agreement with mean-field
results than with PY results. More recently, the same model has been
the subject of investigations related to its universality
class,\cite{JGMC97} to the location of the critical point and the
computation of the coexistence curve,\cite{SY96} to the development
of an integral equation theory that includes the first few terms in
the density expansion of the direct correlation function into the
closure approximation,\cite{YS00} to the (partial) total and direct
correlation functions\cite{FP03}
 through accurate Monte Carlo simulations{, and to the static and
 dynamic behavior near the consolute critical point obtained from
 molecular dynamics simulations. \cite{JY04}}

A theoretical approach that has been extensively used in connection
with non-additive hard-core mixtures is the SPT. Apart from the
papers quoted above, Bergmann\cite{B76b} has considered the SPT in
one dimension and compared it with the exact solution. Tenne and
Bergmann examined the SPT for $d=3$ both for positive
non-additivity\cite{B78a} (where they computed the critical density
and the critical non-additivity) and negative
non-additivity.\cite{B78b} Bearman and Mazo also considered the SPT
for a symmetric binary mixture of non-additive hard-disks\cite{BM88}
and pointed out that the phase transition predicted by Tenne and
Bergmann in Ref.\ \onlinecite{TB79} for negative non-additivity was
spurious. The same authors\cite{MB90} introduced a simpler version
of the SPT for $d=2$ and $d=3$ which is consistent with the SPT of
additive mixtures in the appropriate limit but still presents some
other difficulties. Some of these difficulties were addressed by
Schaink,\cite{S93} who introduced an EOS for a binary mixture valid
for small values of the non-additivity. A comparison of SPT
predictions and simulation data may be found in Ehrenberg \textit{et
al.}\cite{ESH90}

The use of computer simulation, both molecular dynamics (MD) and
Monte Carlo (MC), as well as of the usual integral equation approach
of liquid state theory or the perturbation theory (taking either a
one component system or a binary additive hard-core mixture as the
reference system) have also contributed to the investigation of the
properties of non-additive hard-core mixtures. In the same paper
where they presented the exact solution for the one-dimensional
mixture, Lebowitz and Zomick\cite{LZ71} also gave the exact solution
to the PY equation in $d=1$ and a partial solution to the PY
equation in the three-dimensional case. A mathematical analysis of
these two solutions was later given by Penrose and
Lebowitz.\cite{PL72} Perry and Silbert\cite{PS79} also gave an
approximate solution to the PY equation in $d=3$ which confirmed the
earlier results of Lebowitz and Zomick. For equimolar and symmetric
hard-sphere mixtures with negative non-additivity, Nixon and
Silbert\cite{NS84} solved the PY equation, which they found to
improve its agreement with simulation data as the negative
non-additivity increased. Equimolar symmetric binary mixtures have
been studied by Gazzillo.\cite{G87,G91,G95} He has considered the PY
approximation\cite{G87} and  also other closures (the
Martynov--Sarkisov,\cite{MS83} the
Ballone--Pastore--Galli--Gazzillo\cite{BPGG86} and the modified
Verlet\cite{G91,G95} closures). In Ref.\ \onlinecite{G95} he also
addressed a ternary mixture with negative non-additivity that had
been studied earlier through MD simulation by Schaink,\cite{S94}
while he and his collaborators\cite{BPGG86} were apparently the
first to obtain simulation (MC) data for an asymmetric hard-sphere
binary system. In studying binary non-additive Lennard-Jones
mixtures using the RHNC approximation, Anta and Kahl\cite{AK95}
obtained the non-additive hard-sphere bridge functions by solving
the corresponding PY equation. Lomba \textit{et al.}\cite{LALA96}
used a generalized modified Verlet closure to study fluid-fluid
phase separation in symmetric non-additive hard-sphere mixtures,
obtaining good agreement with their own MC simulation data for the
phase diagram. Kahl \textit{et al.}\cite{KBR96} studied a variety of
symmetric binary mixtures of non-additive hard spheres (both with
positive and negative non-additivity) by solving the
Ornstein--Zernike equation with a modified hypernetted-chain-type
closure. Recently, Sierra and Duda\cite{SR01} considered the PY and
Martynov--Sarkisov closures to study symmetric mixtures of
non-additive hard spheres adsorbed on a disordered hard-sphere
matrix, while Duda \textit{et al.}\cite{DVA03} used MC simulations
to study fluid-fluid phase equilibria and interfacial properties of
non-additive binary hard-sphere mixtures adsorbed in a slit pore.
 The structure and the thermodynamics of non-additive
hard-sphere mixtures under confinement have also been the subject of
a recent study by Pellicane \textit{et al.},\cite{PCWL04} who used
both integral equations and computer simulations.

Melnyck and Sawford\cite{MS75} reported MD simulation data on a
symmetric binary non-additive hard-sphere mixture with positive
non-additivity and using perturbation theory derived an EOS for this
kind of systems which they named MIX1. Such EOS was later extended
to cope with asymmetric mixtures by Schaink and Hoheisel.\cite{SH92}
At about the same time as the Melnyck and Sawford calculations,
Adams and McDonald\cite{AM75} performed MC simulations on binary
symmetric hard-sphere mixtures with negative non-additivity. Later
on, Dickinson\cite{D77} performed MD simulations on two equimolar
non-additive binary hard-disk mixtures. In 1989, Amar\cite{A89}
computed the coexistence curve for the system studied in Ref.\
\onlinecite{MS75} using MC simulation. Hoheisel\cite{H90} studied a
symmetric equimolar binary mixture of non-additive soft spheres with
(high) positive non-additivity through MD and determined the
critical density. Mountain and Harvey\cite{MH91} conducted both MD
and MC simulations on binary mixtures of non-additive soft disks to
study fluid-fluid coexistence. Rovere and Pastore\cite{RP94}
extended the work of Ref.\ \onlinecite{AM75} and obtained the
coexistence curve of an asymmetric binary non-additive hard-sphere
mixture through MC simulation. Extensive MC computations on
symmetric non-additive hard-sphere binary mixtures have been
provided by Jung \textit{et al.},\cite{JJR94a,JJR94b,JJR95} who have
derived from them reasonably accurate (semi-empirical) equations of
state for these systems. Density functional theory has also been
applied\cite{RSLT97} to the computation of the excess free energy of
an equimolar mixture of non-additive hard disks. Finally, recently
Hamad has reported MD calculations for asymmetric non-additive
binary hard-sphere mixtures\cite{H97} and, together with some
coworkers, also for binary hard-disk mixtures.\cite{H99} Fluid-fluid
phase separation in a symmetric mixture of non-additive hard spheres
with positive non-additivity and the phase behavior of non-additive
hard-core mixtures in two dimensions have been recently the subject
of MC simulations by Saija \textit{et al.}\cite{SPG98} and by Saija
and Giaquinta,\cite{SG02} respectively, while G\'o\'zd\'z\cite{G03}
performed MC simulations to derive accurate results for the critical
packing fraction at a few values of the non-additivity parameter in
the case of hard spheres. Casta\~neda-Priego \textit{et
al.}\cite{CRM03} studied depletion interactions in mixtures of
non-additive hard disks, Schmidt\cite{Sch04} generalized the
fundamental measure density functional theory of hard spheres to
binary mixtures of arbitrary positive and moderate negative
non-additivity, and Fantoni and Pastore\cite{FP04} performed
accurate MC simulations to check the local dependency assumption of
the bridge functions of an equimolar non-additive binary hard-sphere
mixture. Fairly recently, Buhot\cite{B04} used a cluster algorithm
to simulate and study phase separation in symmetric binary mixtures
of non-additive hard disks and hard spheres for various (large)
non-additivities including the limiting case of the WR model.

An alternative route to the derivation of the EOS of non-additive
hard-sphere mixtures that does not require neither the SPT,
perturbation theory, the solution of integral equations or
simulation results, relies on the knowledge of virial coefficients
and on the use of exact statistical mechanical relationships. The
so-called $y$-expansion for hard particle fluids introduced by
Barboy and Gelbart\cite{BG79} is a prominent example of this
approach. In the case of non-additive hard-sphere mixtures, the
Barboy--Gelbart EOS involves up to the exact third virial
coefficients, whose analytical expressions are known.\cite{TK55}
 On a different path, Hamad\cite{H96b} has provided a
theory for obtaining mixture properties from pure species
equations of state. In the case of non-additive hard-sphere
mixtures, he invokes exact results pertaining to the contact
values of the radial distribution functions,\cite{H94,H96a,H96c}
as well as the knowledge of the exact second and third virial
coefficients. He has also presented a similar approach for
hard-disk mixtures in Refs.\ \onlinecite{H99,H00}. A noteworthy
aspect of Hamad's proposal is that, due to his use of the
one-component radial distribution function as a starting point,
it is geared essentially towards mixtures not very asymmetric in
size. This proposal has been very recently used in connection
with the development of a perturbation theory for fused sphere
hard-chain fluids.\cite{ASH04}

Recently,\cite{SYH99, SYH02} we have proposed an EOS for a
multicomponent mixture of additive hard-core particles in $d$
dimensions. This proposal shares with Hamad's
approach\cite{H99,H96b,H00} two aspects. On the one hand, it is
expressed in terms of the pure species EOS. And on the other it
starts with a sensible ansatz on the functional form of the contact
values of the radial distribution functions. The aim of this paper
is to complement Hamad's approach in two different veins. The first
one concerns dimensionality. Here we want to derive an EOS for a
non-additive hard-core mixture of an arbitrary number of components
and for any value of $d$. The second one has to do with the fact
that when the non-additivity parameter vanishes we also want to
recover our former proposal\cite{SYH99} for additive multicomponent
hard-core mixtures. Our main concern is to try to keep a reasonable
compromise between the simplicity of the proposal and its ability to
deal also with highly asymmetric mixtures.

The paper is organized as follows. In Sec.\ \ref{sec2} we provide
general expressions for a multicomponent mixture of non-additive
hard-spheres in $d$ dimensions and  some key background material
(third virial coefficients, for which a simple expression for
arbitrary dimensionality is proposed) for the later development. The
exact solution in the case of a one-dimensional binary mixture as
well as other interesting features of this system are presented in
Appendix \ref{appA}. {Section \ref{sec3} contains a brief account of
Hamad's proposal\cite{H96a,H96c,H99} for the contact values of the
radial distribution functions and for the compressibility factor of
the mixture. This is followed in Sec.\ \ref{sec4} by our own
proposal, which shares  with Hamad's  a few features: the
construction of the EOS via the contact values of the radial
distribution functions, the dependence of the latter on the EOS of
the one-component fluid, and the use of the third virial
coefficients.} The results pertaining to special limiting cases are
given in Appendix \ref{appB}. The analysis of the fourth, fifth, and
sixth virial coefficients and of the compressibility factors in one,
two, and three dimensions is carried out in Sec.\ \ref{sec5}. The
paper is closed in Sec.\ \ref{sec6} with further discussion and some
concluding remarks.

\section{Third virial coefficients \label{sec2}}
\subsection{General equations.}
Let us consider an $N$-component mixture of hard spheres in $d$
dimensions. The hard core of the interaction between a sphere of
species $i$ and a sphere of species $j$ is $\sigma_{ij}$. The
diameter of a sphere of species $i$ is $\sigma_{ii}=\sigma_i$. In
general,
$\sigma_{ij}=\frac{1}{2}(\sigma_i+\sigma_j)(1+\Delta_{ij})$, where
$\Delta_{ij}\geq -1$ is a symmetric matrix with zero diagonal
elements ($\Delta_{ii}=0$) that characterizes the degree of
non-additivity of the interactions. In the case of a binary
mixture ($N=2$), the only non-additivity parameter is
$\Delta=\Delta_{12}=\Delta_{21}$. The compressibility factor of
the mixture $Z\equiv p/\rho k_{B}T$, where $\rho$ is the total
number density, $p$ is the pressure, $T$ is the temperature, and
$k_B$ is the Boltzmann constant, can be exactly expressed in
terms of the radial distribution functions at contact $g_{ij}$ as
\beqa
Z(\rho,\{x_k\},\{\sigma_{k\ell}\})&=&1+2^{d-1}v_d\rho\sum_{i,j=1}^{N}x_ix_j
\sigma_{ij}^d\nn\\
&&\times g_{ij}(\rho,\{x_k\},\{\sigma_{k\ell}\}),
\label{1}
\eeqa
where $x_i=\rho_i/\rho$ is the mole fraction of species $i$,
$\rho_i$ is the partial number density of particles of species $i$,
and $v_d=(\pi/4)^{d/2}/\Gamma(1+d/2)$ is the volume of a
$d$-dimensional sphere of unit diameter. Although no general
expression is known for
$g_{ij}(\rho,\{x_k\},\{\sigma_{k\ell}\})\equiv g_{ij}(\rho)$, it can
be expanded in a power series in density as
\begin{equation}
g_{ij}(\rho)=1+v_{d}\rho \sum_{k=1}^{N}x_{k}c_{k;ij}+(v_{d}\rho)^2
\sum_{k,\ell=1}^{N}x_{k}x_\ell c_{k\ell;ij}+{O}(\rho^3).
\label{1M}
\end{equation}
{The coefficients $c_{k;ij}$, $c_{k\ell;ij}$, \ldots are independent
of the composition of the mixture, but they are in general
complicated nonlinear functions of the diameters $\sigma_{ij}$,
$\sigma_{ik}$, $\sigma_{jk}$, $\sigma_{k\ell}$, \ldots Insertion of
the expansion (\ref{1M})}  into Eq.\ (\ref{1}) yields the virial
expansion of $Z$, namely
\begin{eqnarray}
Z(\rho)&=&1+\sum_{n=1}^\infty (v_d\rho)^n \bar{B}_{n+1} \nonumber\\
&=&1+v_{d}\rho \sum_{i,j=1}^{N}\bar{B}_{ij}x_{i}x_{j}+(v_{d}\rho
)^{2}\sum_{i,j,k=1}^{N}\bar{B}_{ijk}x_{i}x_{j}x_{k}\nonumber\\
&&+(v_{d}\rho )^{3}\sum_{i,j,k,\ell=1 }^{N}\bar{B}_{ijk\ell
}x_{i}x_{j}x_{k}x_{\ell }+ {O}(\rho^4). \label{2M}
\end{eqnarray}
Note that, for further convenience, we have introduced the
coefficients $\bar{B}_n\equiv v_d^{-(n-1)} B_n$ where $B_n$ are the
usual virial coefficients. The composition-independent second,
third, and fourth (barred) virial coefficients are given by
\begin{equation} \bar{B}_{ij}=2^{d-1}\sigma_{ij}^d, \label{n1}
\end{equation}
\begin{equation}
\bar{B}_{ijk}=\frac{2^{d-1}}{3}\left(c_{k;ij}\sigma_{ij}^d+
c_{j;ik}\sigma_{ik}^d+c_{i;jk}\sigma_{jk}^d\right), \label{n2}
\end{equation}
\beqa
\bar{B}_{ijk\ell}&=&\frac{2^{d-1}}{6}\left(c_{k\ell;ij}\sigma_{ij}^d+
c_{j\ell;ik}\sigma_{ik}^d+c_{i\ell;jk}\sigma_{jk}^d\right. \nn\\
&& \left.+ c_{jk,i\ell}\sigma_{i\ell}^d+c_{ik,j\ell}\sigma_{j\ell}^d
+c_{ij;k\ell}\sigma_{k\ell}^d\right).
\label{n3}
\eeqa

This connection between the virial coefficients of the mixture and
the $c$'s of the density expansion of the contact values of the
radial distribution functions may be profitably used to devise
sensible approximations.

{For subsequent use in Secs.\ \ref{sec3} and \ref{sec4}, it is
convenient to consider the special case of a one-component fluid
($\sigma_{ij}=\sigma$) of packing fraction $y=v_d\rho\sigma^d$. In
such a case, Eqs.\ (\ref{1}) and (\ref{2M}) become
\beqa
Z_{\text{pure}}(y)&=&1+2^{d-1}y g_{\text{pure}}(y)\nn\\
&=&1+\sum_{n=1}^\infty b_{n+1}y^n,
\label{pure}
\eeqa
where $b_n=\bar{B}_n/\sigma^{(n-1)d}$ are the (reduced) virial
coefficients of the one-component hard-sphere fluid. In particular,
$b_2=2^{d-1}$.}

\subsection{The one-dimensional case.}
It is worth recalling that, as mentioned in the Introduction, in
the case of a \textit{binary} ($N=2$) one-dimensional ($d=1$)
mixture with nearest-neighbor interactions only [which implies
that $2\sigma_{12}\geq \text{max}(\sigma_1,\sigma_2)$], the exact
compressibility factor is known.\cite{PL54,K55,LZ71,CB99,Penrose}
 In Appendix \ref{appA} we provide a
summary of the exact solution as well as some interesting properties
of the same that, to our knowledge, have not been reported before.
In particular, the coefficients $c_{k;ij}$  for $d=1$ are
\begin{equation} c_{1;11}=\sigma_1,\quad
c_{2;11}=2\sigma_{12}-\sigma_1,\quad c_{1;12}=\sigma_1.
\label{H.1}
\end{equation}
The remaining coefficients are obtained from (\ref{H.1}) by the
exchange $1\leftrightarrow 2$.

\subsection{The three-dimensional case.}
\begin{figure}[tbp]
\includegraphics[width=0.8\columnwidth,angle=-90]{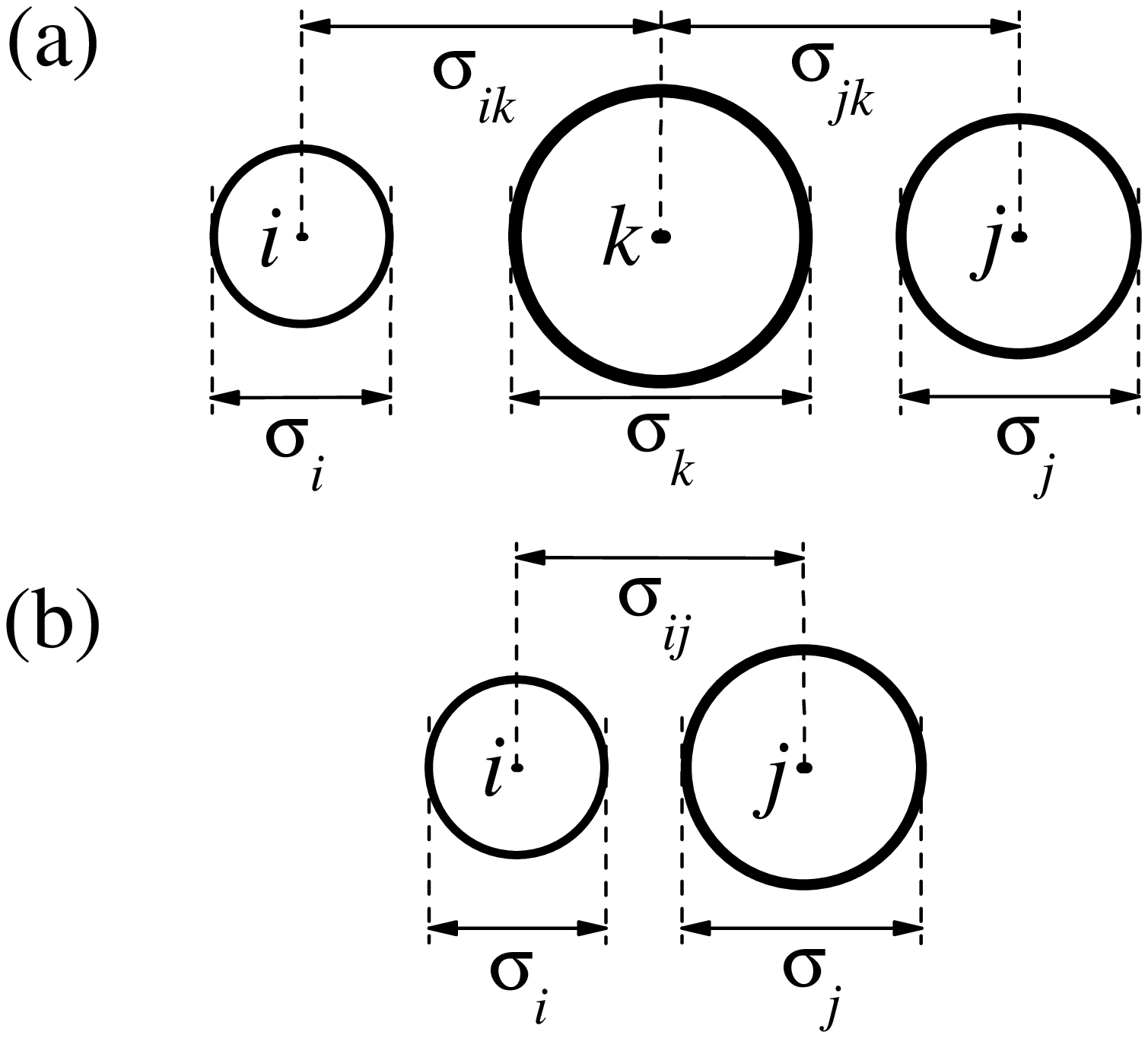}
\caption{(a) Three spheres of species $i$,$j$, and $k$ in an aligned
configuration. The smallest possible separation between spheres $i$
and $j$ is $\sigma_{ik}+\sigma_{jk}$. (b) When sphere $k$ is
removed, the smallest distance between $i$ and $j$ is $\sigma_{ij}$.
Thus $\sigma_{k;ij}=\sigma_{ik}+\sigma_{jk}-\sigma_{ij}$ represents
an effective diameter of sphere $k$ as seen from the point of view
of the pair $ij$. In the sketch we have assumed for simplicity that
the non-additivities are positive. \label{figg}}
\end{figure}

In three dimensions, the first two terms of the exact density
expansion of $g_{ij}$  are known.\cite{H96a} After a few simple
manipulations one may derive from them the result
\begin{equation}
c_{k;ij}=\sigma_{k;ij}^3+\frac{3}{2}\frac{\sigma_{k;ij}^2}{\sigma_{ij}}
\sigma_{i;jk}\sigma_{j;ik} ,
\label{C3bis}
\end{equation}
where
\begin{equation}
\sigma_{k;ij}\equiv\sigma_{ik}+\sigma_{jk}-\sigma_{ij}
\label{diamekk}
\end{equation}
and it is understood that $\sigma_{k;ij}\geq 0$ for all sets $ijk$.
Clearly, $\sigma_{i;ij}=\sigma_i$, $\sigma_{j;ij}=\sigma_j$, {and,
in case of additive hard spheres, $\sigma_{k;ij}=\sigma_k$.} Note
also that the quantities $\sigma_{k;ij}$ may be given a simple
geometrical interpretation. Assume that we have three spheres of
species $i$, $j$, and $k$ aligned in the sequence $ikj$. In such a
case, the distance of closest approach between the centers of
spheres $i$ and $j$ is $\sigma_{ik}+\sigma_{jk}$. If the sphere of
species $k$ were not there, that distance would of course be
$\sigma_{ij}$. Therefore $\sigma_{k;ij}$ as given by Eq.\
(\ref{diamekk}) represents a kind of effective diameter of sphere
$k$, as seen from the point of view of the interaction between
spheres $i$ and $j$. A schematic representation of this
interpretation is provided in Fig.\ \ref{figg}.

\subsection{A generalization to $d$-dimensions.}
It is tempting to extend Eqs.\ (\ref{H.1}) and (\ref{C3bis}) to
$d$ dimensions as
\begin{equation}
c_{k;ij}=\sigma_{k;ij}^d+\left(\frac{b_3}{b_2}-1\right)\frac{\sigma_{k;ij}^{d-1}}{\sigma_{ij}}
\sigma_{i;jk}\sigma_{j;ik}.
\label{n7}
\end{equation}
More specifically, for a binary mixture Eq.\ (\ref{n7}) yields
\beq
\begin{array}{l}
c_{1;11}=\frac{b_3}{b_2}\sigma_1^d,\\
c_{2;11}=(2\sigma_{12}-\sigma_1)^d+
\left(\frac{b_3}{b_2}-1\right)\sigma_1(2\sigma_{12}-\sigma_1)^{d-1},\\
 c_{1;12}={\sigma}_1^d+
\left(\frac{b_3}{b_2}-1\right){(2\sigma_{12}-\sigma_1)\sigma_1^d}/{\sigma_{12}}
.
\end{array}
\label{43c}
\eeq
 Obviously, Eq.\ (\ref{43c})
reduces to Eq.\ (\ref{H.1}) for $d=1$ ($b_2=b_3=1$), while Eq.\
(\ref{n7}) reduces to Eq.\ (\ref{C3bis}) for $d=3$ ($b_2=4$,
$b_3=10$).

All of the above results are restricted to the situation
$\sigma_{k;ij}\geq 0$ for any choice of $i$, $j$, and $k$, {i.e.},
$2\sigma_{12}\geq \text{max}(\sigma_1,\sigma_2)$ in the binary case.
This excludes the possibility of dealing with mixtures with
extremely high negative non-additivity in which one sphere of
species $k$ might `fit in' between two spheres of species $i$ and
$j$ in contact. Since for $d=3$ and $N=2$ the coefficients
$c_{k;ij}$ are also known for such mixtures,\cite{H96b} we may
extend our proposal to deal with these cases. If $N=2$, one has
specifically
\beq
\begin{array}{l}
 c_{1;11}=\frac{b_3}{b_2}\sigma_1^d,\quad
 c_{2;11}=\widehat{\sigma}_{2}^d+
\left(\frac{b_3}{b_2}-1\right)\sigma_1\widehat{\sigma}_{2}^{d-1},\\
 c_{1;12}=(2\sigma_{12}-\widehat{\sigma}_2)^d+
\left(\frac{b_3}{b_2}-1\right){\widehat{\sigma}_2\sigma_1^d}/{\sigma_{12}}
,
\end{array}
\label{negative}
\eeq
where we have defined
\begin{equation}
\widehat{\sigma}_2=\text{max}\left(2\sigma_{12}-\sigma_1,0\right).
\label{negative2}
\end{equation}
With such an extension, we recover the exact values of $c_{k;ij}$
for a binary mixture of hard spheres ($d=3$), even if $\sigma_1> 2
\sigma_{12}$ or $\sigma_2> 2 \sigma_{12}$. We emphasize that Eqs.\
(\ref{n7})--(\ref{negative2}) for $d \neq 1$ and $d \neq 3$ are new.

\subsection{The two-dimensional case}
While Eq.\ (\ref{negative}) is exact for $d=1$ and $d=3$, it is only
approximate for $d=2$. For that dimensionality, the exact result has
been derived by Al-Naafa \textit{et al.}\cite{H99} After some
algebra (and the correction of some typos), the coefficients
$c_{k;ij}$ can be written as
\begin{equation}
\begin{array}{l}
 c_{1;11}=\frac{b_3}{2}\sigma_1^2,\quad
c_{2;11}=\frac{b_3}{2}\sigma_1^2
F\left({\sigma_{12}}/{\sigma_1}\right),\\
 c_{1;12}=\frac{b_3}{2}\sigma_1^2
G\left({\sigma_{12}}/{\sigma_1}\right) ,
\end{array}
\label{H.2}
\end{equation} where
$b_3=\frac{16}{3}-\frac{4\sqrt{3}}{\pi}\simeq 3.1280$ and the
functions $F(s)$ and $G(s)$ are given by
\begin{equation} F(s)=
\left\{
\begin{array}{lr}
\frac{4}{\pi b_3}\left(4s^2\cos^{-1} \frac{1}{2s}-\sqrt{4s^2-1}\right),& s\geq \frac{1}{2},\\[0.1cm]
0,& 0\leq s\leq \frac{1}{2},
\end{array} \right.
\label{H.3}
\end{equation}
\begin{equation}
G(s)= \left\{
\begin{array}{lr}
\frac{4}{\pi b_3}\left[2\pi s^2-2\left(2s^2-1\right)\cos^{-1}
\frac{1}{2s} \right.&\\
\left. -\sqrt{4s^2-1}\right], & s\geq \frac{1}{2},
\\[0.1cm]
 \frac{8}{b_3}s^2,&0\leq s\leq \frac{1}{2}.
\end{array}
\right.
\label{H.4}
\end{equation}
Some special values of $F(s)$ and $G(s)$ are
\begin{equation}
F\left(1\right)=G\left(1\right)=1,
\label{H.6}
\end{equation}
\begin{equation}
F\left({1}/{2}\right)=0,\quad G\left({1}/{2}\right)=\frac{2}{b_3},
\label{H.5}
\end{equation}
\begin{equation}
\lim_{s\to\infty}s^{-2}F(s)=\frac{8}{b_3},\quad
\lim_{s\to\infty}G(s)=\frac{4}{b_3}.
\label{H.7}
\end{equation}
For a symmetric mixture ($\sigma_1=\sigma_2$), the value
$s=\sigma_{12}/\sigma_1=1$ corresponds to the one-component case,
$s=\sigma_{12}/\sigma_1=\frac{1}{2}$ corresponds to the threshold
value of negative non-additivity ({i.e.},
$2\sigma_{12}=\sigma_1=\sigma_2$ or $\Delta=-\frac{1}{2}$), and
the limit $s=\sigma_{12}/\sigma_1\to \infty $ represents an
infinitely large positive non-additivity (WR model).

\begin{figure}[tbp]
\includegraphics[width=0.9\columnwidth]{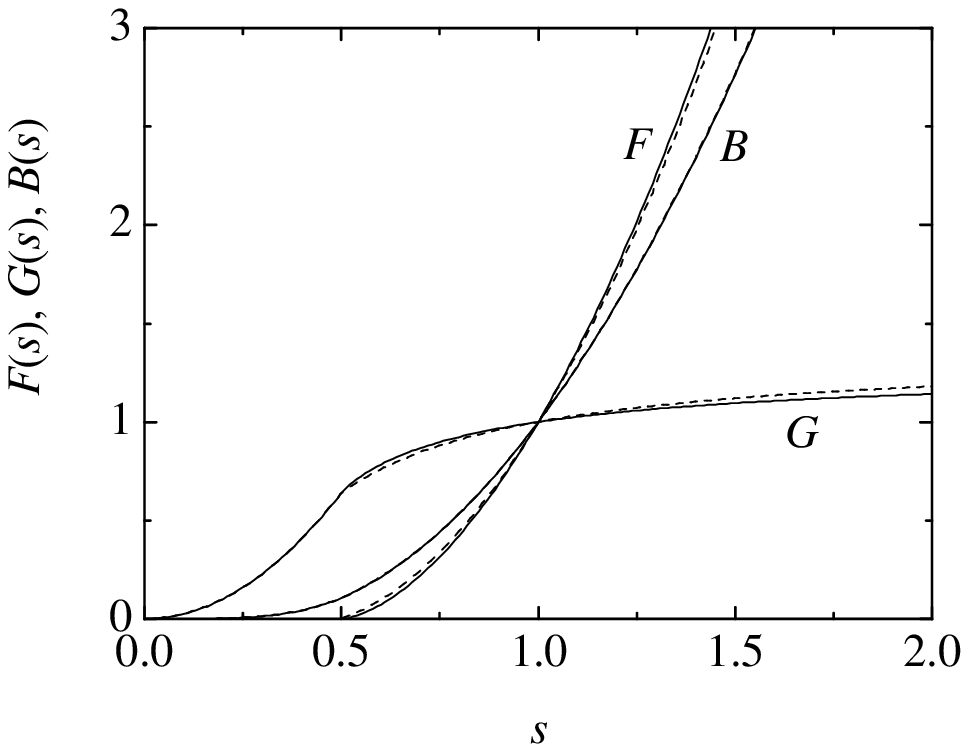}
\caption{Plot of the functions $F(s)$, $G(s)$, and $B(s)$. The solid
lines are the exact functions (\protect\ref{H.3}),
(\protect\ref{H.4}), and (\protect\ref{H.13}), while the dashed
lines are our approximations (\protect\ref{H.8.1}),
(\protect\ref{H.8.2}), and (\protect\ref{H.13.2}). Note that $B(s)$
and $B_{\text{app}}(s)$ are practically indistinguishable.
\label{FyG}}
\end{figure}

Equation (\ref{negative}) with $d=2$ can be recast into the form
(\ref{H.2}), except that the functions $F(s)$ and $G(s)$ are
approximated by
\begin{equation}
F_{\text{app}}(s)= \left\{
\begin{array}{lr}
\frac{1}{b_3}\left(2s-1\right)\left(4s+b_3-4\right),& s\geq \frac{1}{2},\\
0,& 0\leq s\leq \frac{1}{2},
\end{array}
\right.
\label{H.8.1}
\end{equation}
\begin{equation}
G_{\text{app}}(s)= \left\{
\begin{array}{lr}
\frac{2}{b_3}\left(b_3-1-\frac{b_3-2}{2s}\right),& s\geq \frac{1}{2},\\
 \frac{8}{b_3}s^2,&0\leq s\leq \frac{1}{2}.
\end{array}
\right.
\label{H.8.2}
\end{equation}
This approximation verifies the properties
(\ref{H.5})--(\ref{H.7}), except that now
$\lim_{s\to\infty}G_{\text{app}}(s)=2(b_3-1)/b_3$, which is about
6\% higher than the exact value. Figure \ref{FyG} shows that
Eqs.\ (\ref{H.8.1}) and (\ref{H.8.2}) constitute an excellent
approximation to the exact expressions (\ref{H.3}) and
(\ref{H.4}), especially for small or moderate values of $s$.

The third virial coefficients $\bar{B}_{ijk}$ for a
two-dimensional binary mixture as given by Eq.\ (\ref{n2}) may be
cast into the form
\begin{equation}
\bar{B}_{111}=2c_{1;11}\sigma_1^2=b_3\sigma_1^4 ,
\label{H.9}
\end{equation}
\begin{eqnarray}
\bar{B}_{112}&=&\frac{2}{3}\left(c_{2;11}\sigma_1^2+2c_{1;12}\sigma_{12}^2\right)\nonumber\\
&=&b_3\sigma_1^4 B\left({\sigma_{12}}/{\sigma_1}\right) ,
\label{H.10}
\end{eqnarray}
and similar expressions for $\bar{B}_{222}$ and $\bar{B}_{221}$,
obtained from the former by the exchange of indices $1$ and $2$.
Here,
\begin{eqnarray}
B(s)&\equiv& \frac{1}{3}F(s)+\frac{2}{3}s^2 G(s) \nonumber \\
&=&\left\{
\begin{array}{lr}
\frac{4}{3\pi b_3}\left[4\pi s^4-8s^2\left(s^2-1\right)\cos^{-1}
\frac{1}{2s}\right.&\\
\left.-\left(2s^2+1\right) \sqrt{4s^2-1}\right], & s\geq
\frac{1}{2},
\\[0.1cm]
\frac{16}{3b_3}s^4,& 0\leq s\leq \frac{1}{2},
\end{array}
\right.\nn\\
\label{H.13}
\end{eqnarray}
Using Eqs.\ (\ref{H.8.1}) and (\ref{H.8.2}) our approximation yields
for $B_{\text{app}}(s)$ the polynomial
\begin{equation} B_{\text{app}}(s)= \left\{
\begin{array}{lr}
\frac{1}{3b_3}\left[4(s-1)^2+b_3\left(4s^2-1\right)\right],& s\geq \frac{1}{2},\\
\frac{16}{3b_3}s^4,& 0\leq s\leq \frac{1}{2}.
\end{array}
\right.
\label{H.13.2}
\end{equation}
As also seen in Fig.\
\ref{FyG}, Eq.\ (\ref{H.13.2}) is practically indistinguishable from
the exact $B(s)$, so that the (small) discrepancies in
$F_{\text{app}}(s)$ and $G_{\text{app}}(s)$ with respect to the
actual $F(s)$ and $G(s)$ almost entirely compensate. Therefore, it
seems that it is not unreasonable to use Eqs.\
(\ref{n7})--(\ref{negative2}) for all $d$.

\section{{Hamad's proposal for the equation of state} \label{sec3}}

Our goal is to derive an (approximate) EOS for a multicomponent
mixture of $d$-dimensional non-additive hard spheres. Clearly, this
may be achieved if values for the $g_{ij}$ are provided. {But before
we engage in this task, let us recall in this section a previous
simple proposal by Hamad.}

Hamad\cite{H96a,H96c,H99} has proposed a simple and accurate
approximation for the contact values of the radial distribution
functions which takes the same form in both $d=2$ and $d=3$.
Generalized to arbitrary dimensionality $d$ and in the notation
of this paper it reads
\begin{equation}
g_{ij}^{\text{H}}(\rho)=g_{\text{pure}}\left(\eta X_{ij}\right),
\quad X_{ij}=\frac{b_2}{b_3}\frac{\sum_{k}x_kc_{k;ij}}{\langle
\sigma^d\rangle}.
\label{H.19}
\end{equation}
{Here, $\eta\equiv v_d\rho \langle\sigma^d\rangle$ is the packing
fraction of the mixture, with
${\langle\sigma^m\rangle}=\sum_{i=1}^{N}x_i\sigma_i^m$.}

By construction, the aproximation (\ref{H.19}) is correct to
first order in the density (third virial coefficient). Inserting
the approximation (\ref{H.19}) into Eq.\ (\ref{1}), we obtain the
(generalized) $d$-dimensional Hamad EOS
\begin{eqnarray}
Z^{\text{H}}(\rho)&=&1+\frac{2^{d-1}\eta}
{\langle\sigma^d\rangle}\sum_{i,j}x_i
x_j\sigma_{ij}^dg_{\text{pure}}\left(\eta X_{ij}\right)\nonumber\\
&=&1+\sum_{i,j}\frac{x_i
x_j\sigma_{ij}^d}{\langle\sigma^d\rangle}\left[
\frac{Z_{\text{pure}}\left(\eta X_{ij}\right)-1}{X_{ij}} \right] .
\label{H.22}
\end{eqnarray}

So far, the one-component function $Z_{\text{pure}}(\etap)$ remains
free. It should be emphasized that, except for $d=2$ and $d=3$, the
EOS given by Eq.\ (\ref{H.22}) has been {neither} introduced nor
used before.

The Helmholtz free energy  per particle of a mixture, $a(\rho)$, is
given by
\beq
\frac{a(\rho)}{ k_{B}T}= -1+\sum_{i=1}^{N}x_{i}\ln
\left( \rho_{i}\lambda_{i}^d\right)+\int_{0}^{\rho }\frac{d\rho
\prime}{\rho \prime}\left[ Z(\rho \prime )-1\right],
\label{FEN}
\eeq
where $\lambda_{i}$ is the thermal de Broglie wavelength of
species $i$. According to Hamad's approximation (\ref{H.22}),
\beq
\frac{a^{\text{H}}(\rho)}{ k_{B}T}= -1+\sum_{i}x_{i}\ln
\left(
\rho_{i}\lambda_{i}^d\right)+\sum_{i,j}\frac{x_ix_j\sigma_{ij}^d}{\langle\sigma^d\rangle
X_{ij}}\frac{a_{\text{pure}}^{\text{ex}}\left(\eta
X_{ij}\right)}{k_BT},
\label{FEN-H}
\eeq where
$a_{\text{pure}}^{\text{ex}}(\etap)$ is the excess Helmholtz free
energy  per particle of the pure fluid.

\section{Our proposal \label{sec4}}
In 1999 we proposed an EOS for a multicomponent mixture of
\textit{additive} hard spheres in $d$-dimensions\cite{SYH99} which
was based on an ansatz related to the contact values of the radial
distribution functions. One may express this ansatz as
\begin{equation}
g_{ij}^{\text{SYH}}(\rho)= \frac{1}{1-\eta}
+\left[g_{\text{pure}}(\eta) -\frac{1}{1-\eta} \right] z_{ij},
\label{C1bis}
\end{equation}
where
\begin{equation}
z_{ij}=\frac{\langle \sigma^{d-1}\rangle\sigma_i\sigma_j} {\langle
\sigma^d\rangle \sigma_{ij}}\quad \text{(additive spheres)}
\label{C1bis2}
\end{equation}
is a parameter that is independent of density but depends on the
composition and diameters of the mixture.

The idea is now to generalize the ansatz given by Eq.\ (\ref{C1bis})
to the non-additive case. As the simplest possible extension, we
keep the structure of Eq.\ (\ref{C1bis}) but determine the
parameters $z_{ij}$ as to reproduce Eq.\ (\ref{1M}) to first order
in the density. The result is {readily found to be}
\beq
z_{ij}=\left(\frac{b_3}{b_2}-1\right)^{-1}\left(\frac{\sum_k
x_kc_{k;ij}}{\langle\sigma^d\rangle}-1\right).
\label{new1bis}
\eeq
The following relationship between $z_{ij}$ and $X_{ij}$ exists:
\begin{equation}
z_{ij}=\frac{b_3X_{ij}-b_2}{b_3-b_2},\quad X_{ij}=z_{ij}+
\frac{b_2}{b_3}\left(1-z_{ij}\right).
\label{zX}
\end{equation}

{The ansatz (\ref{C1bis}) supplemented by Eq.\ (\ref{new1bis})  is,
by construction, accurate for densities low enough as to justify the
linear approximation $g_{ij}\approx 1+v_d\rho\sum_k x_k c_{k;ij}$.
On the other hand, the limitations of this truncated expansion for
moderate and large densities are compensated by the use of
$g_{\text{pure}}$. Of course, $g_{ij}=g_{\text{pure}}$ in the
special case where all the diameters are identical
($\sigma_{k\ell}=\sigma$), since then $c_{k;ij}=(b_3/b_2)\sigma^d$
and $z_{ij}=1$. All these comments apply to Hamad's prescription
(\ref{H.19}) as well. On the other hand,  Eq.\ (\ref{C1bis}) is
consistent, but Eq.\ (\ref{H.19}) is not,  with the case of an
additive mixture in which one of the species, say $i=1$, is made of
point particles, so that $g_{11}=(1-\eta)^{-1}$.}

When Eqs.\ (\ref{C1bis}) and (\ref{new1bis}) are inserted into
Eq.\ (\ref{1}) one gets
\beqa
Z^{\text{SYH}}(\rho)&=&1+\frac{\eta}{1-\eta}\frac{b_3\langle\sigma^d\rangle
\bar{B}_2-b_2 \bar{B}_3}{(b_3-b_2)\langle\sigma^d\rangle^2}\nn\\
&&+
\left[Z_{\text{pure}}(\eta)-1\right]\frac{\bar{B}_3-\langle\sigma^d\rangle
\bar{B}_2}{(b_3-b_2)\langle\sigma^d\rangle^2}.
\label{new2}
\eeqa

Equation ({\ref{new2}) is the main result of this paper. As in Eq.\
(\ref{H.22}), the EOS of the mixture is expressed in terms of that
of the one-component system. On the other hand, the density
dependence in the EOS (\ref{new2}) is simpler: $Z(\rho)-1$ is
expressed as a linear combination of $\eta/(1-\eta)$ and
$Z_{\text{pure}}(\eta)-1$, with coefficients such that the second
and third virial coefficients are reproduced. {Again, Eq.\
(\ref{new2}) is accurate for sufficiently low densities, while the
limitations of the truncated expansion for moderate and large
densities are compensated by the use of the EOS of the pure fluid.}

In the approximation (\ref{new2}), the Helmholtz free energy per
particle is
\begin{eqnarray}
\frac{a^{\text{SYH}}(\rho)}{ k_{B}T}&=&  -1+\sum_{i}x_{i}\ln \left(
\rho_{i}\lambda_{i}^d\right)-\ln(1-\eta)\nn\\
&& \times\frac{b_3\langle\sigma^d\rangle \bar{B}_2-b_2
\bar{B}_3}{(b_3-b_2)\langle\sigma^d\rangle^2}+
\frac{a_{\text{pure}}^{\text{ex}}(\eta)}{
k_BT}\frac{\bar{B}_3-\langle\sigma^d\rangle
\bar{B}_2}{(b_3-b_2)\langle\sigma^d\rangle^2}.
\nn\\
\label{FEN-SYH}
\end{eqnarray}

In principle, to compute $\bar{B}_3$, one should use the exact
coefficients $c_{k;ij}$. However, since to the best of our knowledge
they are only known for $d \leq 3$ and we want our proposal to be
explicit for any $d$, we can make use of our approximation for them,
Eq.\ (\ref{n7}). Therefore, with this proviso we get
\beqa
z_{ij}
&=&\left(\frac{b_3}{b_2}-1\right)^{-1}\left(\frac{\sum_k
x_k\sigma_{k;ij}^d}{\langle\sigma^d\rangle}-1\right)\nn\\ &&+
\frac{\sum_k x_k
\sigma_{k;ij}^{d-1}\sigma_{i;jk}\sigma_{j;ik}}{\langle\sigma^d\rangle\sigma_{ij}}
. \label{new1}
\eeqa
In the additive case ($\sigma_{k;ij}\to \sigma_k$), Eq.\
(\ref{new1}) reduces to Eq.\ (\ref{C1bis2}). Note that both for
$d=1$ and $d=3$ there is no difference in the resulting
compressibility factor because Eq.\ (\ref{n7}) yields the exact
result. On the other hand, for other $d$, use of Eq.\ (\ref{new1})
also leads to Eq.\ ({\ref{new2}), but with an approximate rather
than the exact value for the third virial coefficient.

\section{Results \label{sec5}}
Once we have derived our approximation for the EOS of the
mixture, Eq.\ (\ref{new2}), it is interesting to examine its
performance. And since Hamad has carried out a comparison between
his proposal, Eq.\ (\ref{H.22}), and previous
ones,\cite{H96b,H96a,H96c,H99}
 finding in general that it performs better,
we will concentrate here on comparing the results obtained either
through Hamad's prescription or through ours (in this regard see
also Appendix \ref{appB}). Such comparison seems in order in view of
the fact that both proposals share many aspects such as the
construction of the EOS via the contact values of the radial
distribution functions, its dependence on the EOS of the
one-component fluid (more specifically on $Z_{\text{pure}}$, that
remains to be chosen freely) and the use of the third virial
coefficients. Also, and although Hamad's proposal is specific for
$d=2$ and $d=3$ and we have extended it to arbitrary $d$, they
maintain the same form in every dimensionality. Specifically, we
will focus on the fourth and higher virial coefficients and on the
compressibility factor. To our knowledge, and with the exception of
the one-dimensional case, in which they are known exactly, values of
the former are rather scarce\cite{SFG98,VM03} and refer exclusively
to non-additive hard spheres ($d=3$).

\subsection{Fourth and higher virial coefficients }
\begin{figure}[tbp]
\includegraphics[width=0.9\columnwidth]{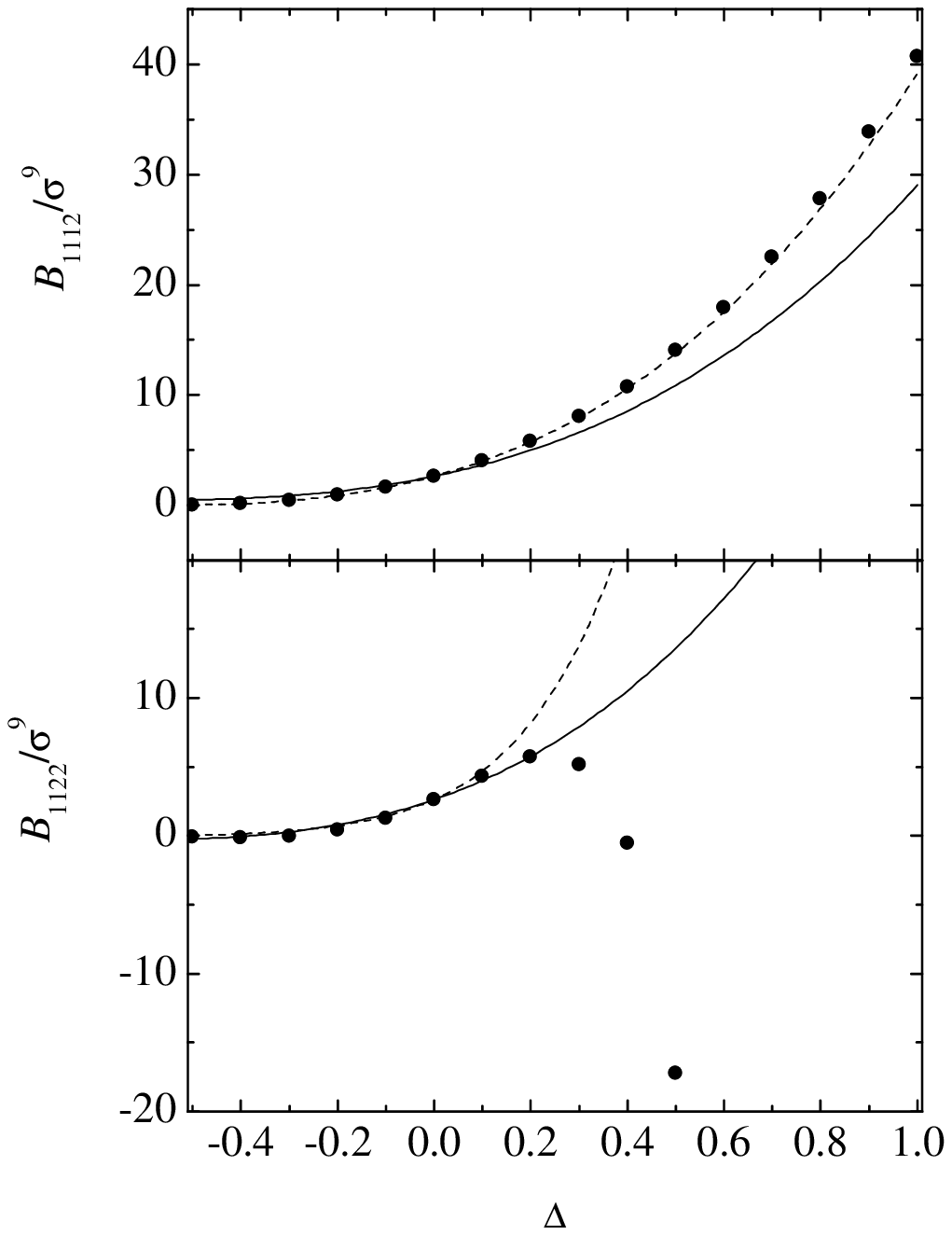}
\caption{Plots of ${B}_{1112}/\sigma^9=(\pi/6)^3
\bar{B}_{1112}/\sigma^9$  and ${B}_{1122}/\sigma^9=(\pi/6)^3
\bar{B}_{1122}/\sigma^9$ versus $\Delta$ for a symmetric
three-dimensional binary mixture. Circles: exact
values;\protect\cite{SFG98} solid lines: Eqs.\ (\protect\ref{5.4})
and (\protect\ref{5.5}) (present approach); dashed lines: Eqs.\
(\protect\ref{5.6}) and (\protect\ref{5.7}) (Hamad's result).
\label{figB1112}}
\end{figure}
\begin{figure}[tbp]
\includegraphics[width=0.9\columnwidth]{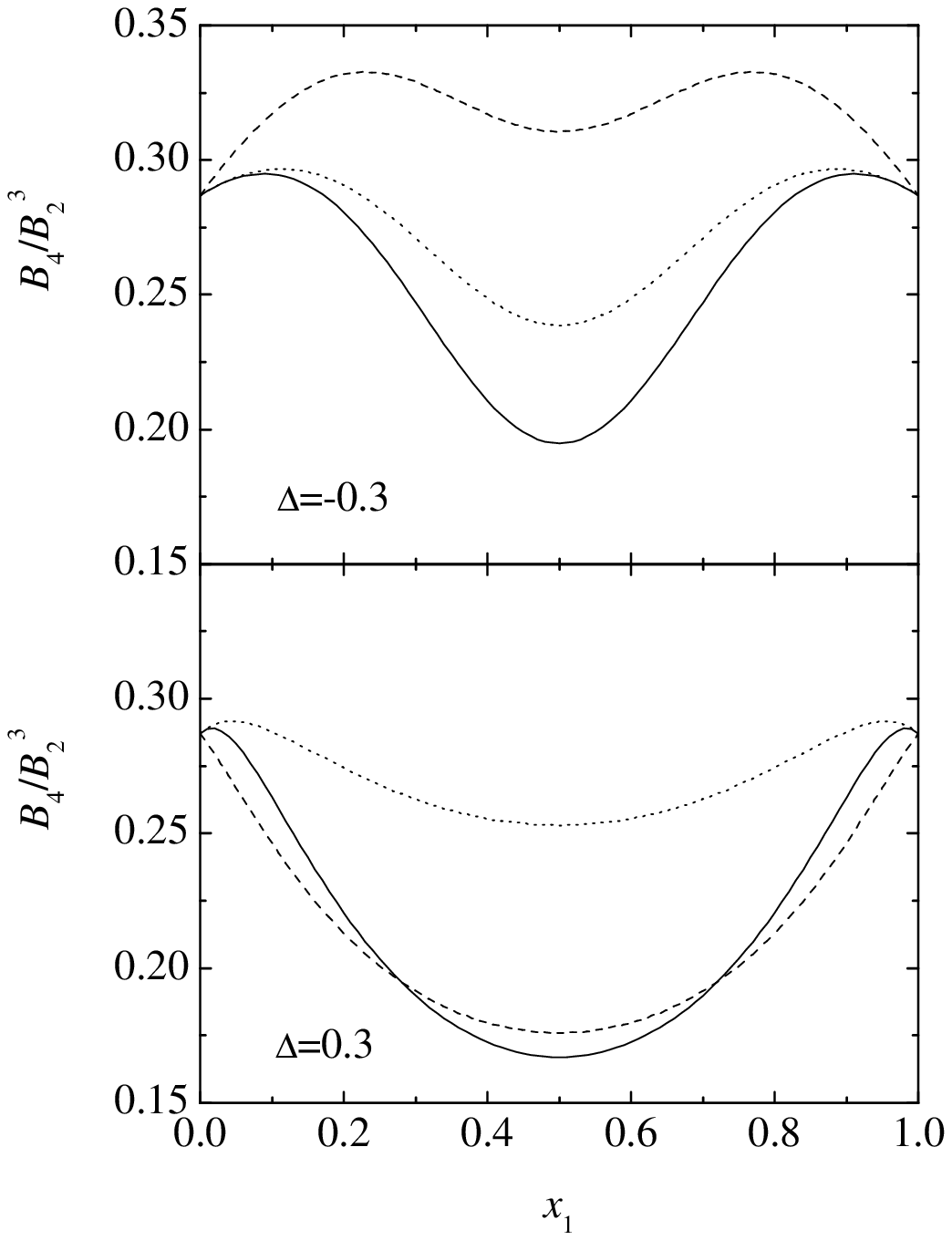}
\caption{Plot of $B_{4}/B_2^3$ versus $x_1$ for a symmetric
three-dimensional binary mixture with $\Delta=-0.3$ and
$\Delta=0.3$. Solid lines: exact values;\protect\cite{SFG98} dashed
line:  Eq.\ (\protect\ref{5.1}) (present approach); dotted line:
Eq.\ (\protect\ref{Hkk5.1}) (Hamad's result).
\label{B4plus}}
\end{figure}
\begin{figure}[tbp]
\includegraphics[width=0.9\columnwidth]{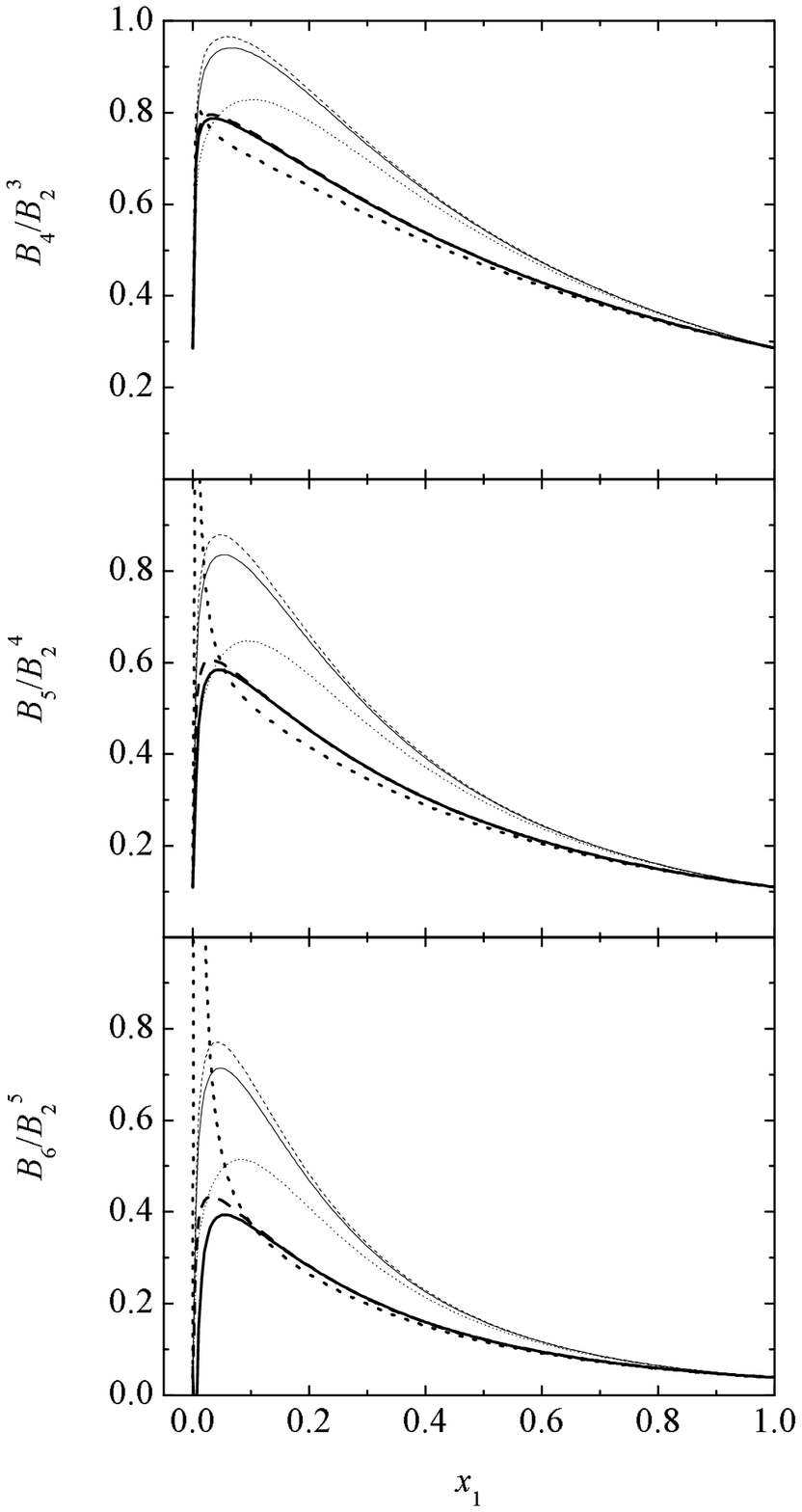}
\caption{Plot of ${B_{n}}/{B_2^{n-1}}$ versus $x_1$ ($n=4,5,6$) for
an asymmetric three-dimensional binary mixture with
${\sigma_2}/{\sigma_1}=0.1$ and $\Delta=0$ (thin lines) and
$\Delta=0.1$ (thick lines). Solid lines: exact
values;\protect\cite{VM03} dashed lines:  Eq.\ (\protect\ref{5.1})
(present approach); dotted lines:  Eq.\ (\protect\ref{Hkk5.1})
(Hamad's result). \label{figMasters}}
\end{figure}
From Eq.\ (\ref{new2}) it is easy to get an approximate expression
for the $n$th virial coefficient:
\begin{equation}
\bar{B}_n^{\text{SYH}}=\frac{b_n-b_2}{b_3-b_2}\langle\sigma^d\rangle^{n-3}
\bar{B}_3-\frac{b_n-b_3}{b_3-b_2}\langle\sigma^d\rangle^{n-2}
\bar{B}_2.
 \label{5.1}
\end{equation}
In particular, the composition independent fourth virial
coefficients are given by
\begin{eqnarray}
\bar{B}_{ijk\ell}^{\text{SYH}}&=&\frac{b_4-b_2}{4(b_3-b_2)}\left(\sigma_i^d
\bar{B}_{jk\ell}+\sigma_j^d \bar{B}_{ik\ell}+\sigma_k^d
\bar{B}_{ij\ell}+\sigma_\ell^d \bar{B}_{ijk}\right)\nonumber\\
&&-\frac{b_4-b_3}{6(b_3-b_2)}\left(\sigma_i^d \sigma_j^d
\bar{B}_{k\ell}+\sigma_i^d \sigma_k^d \bar{B}_{j\ell}+\sigma_i^d
\sigma_\ell^d \bar{B}_{jk}\right.\nn\\
&&\left.+\sigma_j^d \sigma_k^d \bar{B}_{i\ell}+\sigma_j^d
\sigma_\ell^d \bar{B}_{ik}+\sigma_k^d \sigma_\ell^d
\bar{B}_{ij}\right).
\label{5.2}
\end{eqnarray}
In the case of Hamad's approximation, Eq.\ (\ref{H.22}), one has
\begin{equation}
\bar{B}_n^{\text{H}}=b_n\langle\sigma^d\rangle^{n-2}\sum_{i,j}x_i
x_j\sigma_{ij}^d X_{ij}^{n-2},
\label{Hkk5.1}
\end{equation}
\beqa
\bar{B}_{ijk\ell}^{\text{H}}&=&\frac{b_4b_2^2}{6b_3^2}\left(\sigma_{ij}^d
c_{k;ij}c_{\ell;ij}+\sigma_{ik}^d
c_{j;ik}c_{\ell;ik}+\sigma_{i\ell}^d
c_{j;i\ell}c_{k;i\ell}\right.\nn\\
&&\left.+\sigma_{jk}^d c_{i;jk}c_{\ell;jk}+\sigma_{j\ell}^d
c_{i;j\ell}c_{k;j\ell}+\sigma_{k\ell}^d
c_{i;k\ell}c_{j;k\ell}\right).\nn\\
\label{5.3}
\eeqa}
In the special case of  binary and symmetric
[$\sigma_1=\sigma_2=\sigma$, $\sigma_{12}=\sigma(1+\Delta)$]
three-dimensional mixtures, Eqs.\ (\ref{5.2}) and (\ref{5.3}) yield
\beqa
\bar{B}_{1112}^{\text{SYH}}/\sigma^9&=&b_4\left(1+4\Delta+\frac{11}{2}\Delta^2+
\frac{7}{3}\Delta^3\right)\nn\\
&&-\Delta\left(10+16\Delta+\frac{22}{3}\Delta^2\right),
\label{5.4}
\eeqa
\beqa
\bar{B}_{1122}^{\text{SYH}}/\sigma^9&=&b_4\left(1+\frac{48}{9}\Delta+
\frac{22}{3}\Delta^2+\frac{28}{9}\Delta^3\right)\nn\\
&&-\frac{8}{9}\Delta \left(15+24\Delta+11\Delta^2\right),
\label{5.5}
\eeqa
\begin{equation}
\bar{B}_{1112}^{\text{H}}/\sigma^9=b_4\left(1+\frac{9}{2}\Delta+
\frac{162}{25}\Delta^2+\frac{144}{50}\Delta^3\right), \label{5.6}
\end{equation}
\beqa
\bar{B}_{1122}^{\text{H}}/\sigma^9&=&b_4\left(1+6\Delta+\frac{408}{25}
\Delta^2+\frac{672}{25}\Delta^3\right.\nn\\
&&\left.+\frac{688}{25}\Delta^4
+\frac{384}{25}\Delta^5+\frac{256}{75}\Delta^6\right),
\label{5.7}
\eeqa where $b_4=18.36477$ and we have assumed that $\Delta\geq
-\frac{1}{2}$. The two coefficients $\bar{B}_{1112}$ and
$\bar{B}_{1122}$ have been evaluated numerically by Saija \textit{et
al.}\cite{SFG98} Figure \ref{figB1112} compares the numerical data
for $\bar{B}_{1112}$ and $\bar{B}_{1122}$ with the approximations
(\ref{5.4})--(\ref{5.7}). We observe that Hamad's approximation for
$\bar{B}_{1112}$ gives an excellent agreement, while ours is only
qualitatively correct. On the other hand, for $\bar{B}_{1122}$ both
approximations are inaccurate for large positive non-additivities.
In any case, $\bar{B}_{1122}^{\text{SYH}}$ is slightly better than
$\bar{B}_{1122}^{\text{H}}$ for $0<\Delta\lesssim 0.3$.

Figure \ref{B4plus} shows, also for a symmetric binary mixture of
non-additive hard spheres, $B_4/B_2^3$ as a function of the mole
fraction $x_1$ for $\Delta=-0.3$ and $\Delta=0.3$, and the
corresponding simulation results. We observe that Hamad's
approximation is better for $\Delta=-0.3$, while ours is better for
$\Delta=0.3$.

As far as we know, the only report of virial coefficients beyond
the third for the case of an asymmetric non-additive hard-sphere
mixture is due to Vlasov and Masters.\cite{VM03} They have
computed up to the sixth virial coefficient for a binary mixture
of non-additive hard spheres of size ratio $0.1$ and a positive
non-additivity $\Delta=0.1$, and up to the seventh virial
coefficient for a binary (additive) hard-sphere mixture of the
same size ratio. In Fig.\ \ref{figMasters} we present a
comparison of the results for the composition dependence of the
ratio of virial coefficients $B_{n}/B_{2}^{n-1}$ ($n=4,5,6$) in
the case of a binary mixture of size ratio
$\sigma_{2}/\sigma_{1}=0.1$ and two non-additivities ($\Delta=0,
0.1$) given by Vlasov and Masters\cite{VM03} with the results
that follow  from Hamad's prescription and from our proposal. The
overall superiority of our proposal in this case is apparent and
more noticeable for the positive non-additivity and when $n$
increases. Nevertheless, the negative values of the sixth virial
coefficient for the small region around $x = 0$ that are obtained
with the simulation, not shown in the figure, are not captured by
either proposal.

\subsection{ Compressibility factor}
\begin{figure}[tbp]
\includegraphics[width=0.9\columnwidth]{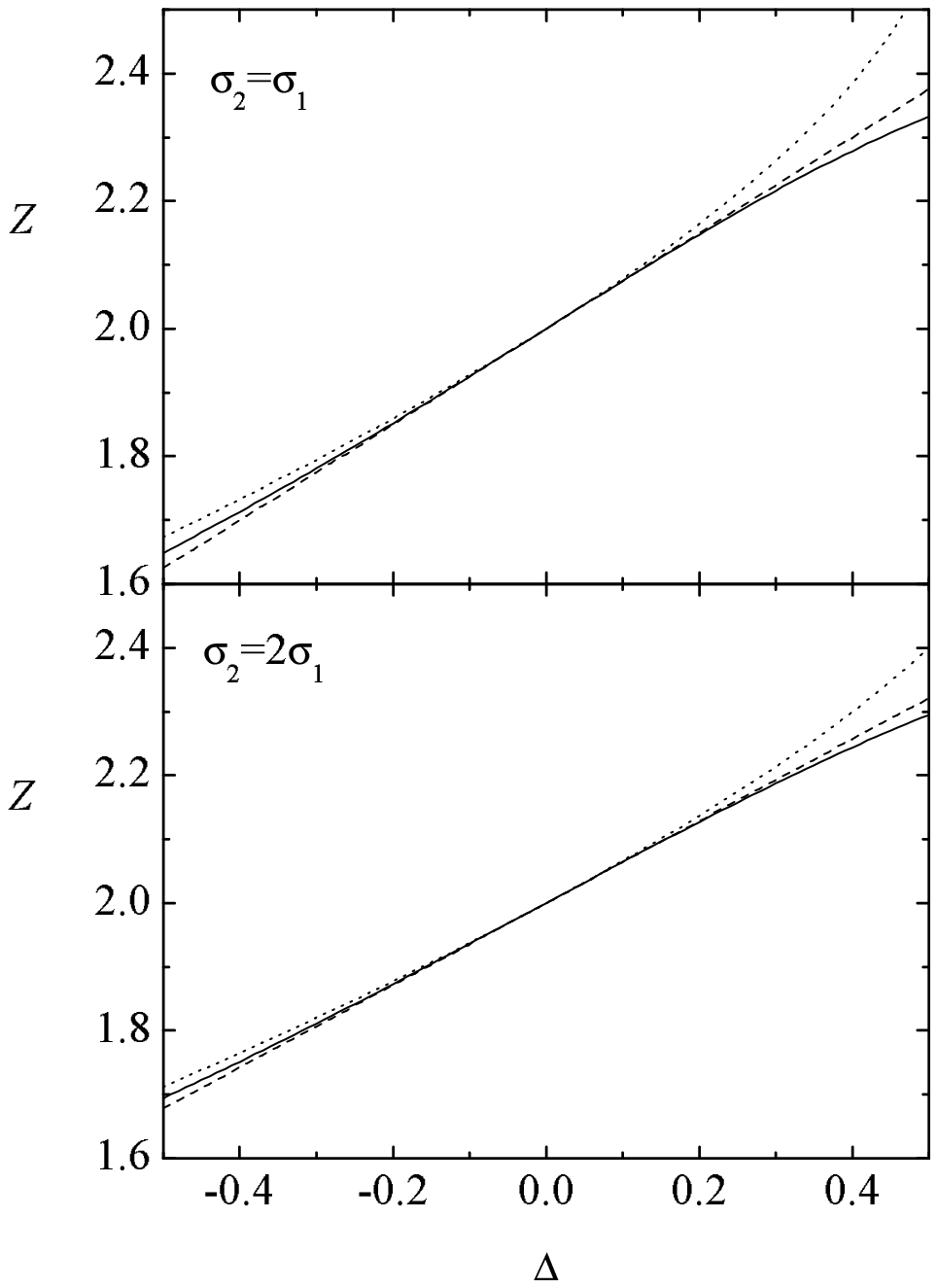}
\caption{Compressibility factor $Z$ as a function of $\Delta$ for a
symmetric mixture of non-additive hard rods with $x_1=0.25$ at a
packing fraction $\eta=0.5$ (upper panel) and for an asymmetric
mixture with $x_1=0.25$ and $\sigma_2/\sigma_1=2$ at $\eta=0.5$
(lower panel). Solid lines: exact; dashed lines: Eq.\
(\protect\ref{B.1}) (present approach); dotted lines: Eq.\
(\protect\ref{B.3}) (Hamad's result).\label{fig1D}}
\end{figure}

Apart from the virial coefficients, the most important
tests concern the compressibility factor itself. In view of the
big number of parameters in these systems, one has to make a
judicious choice such that the main features of the results may
be illustrated. In this subsection we provide a representative set
of data for different dimensionalities that will hopefully cater
for the above requirement.

\subsubsection{Rods ($d=1$) \label{rods}}
In the case $d=1$, one has $Z_{\text{pure}}(\etap)=1/(1-\etap)$
and $b_n=1$ for all $n$, so that our proposal (\ref{new2}) is
ill-defined. To save that singularity and with the aim of
preserving the scaling property of the exact solution (see
Appendix \ref{appA}), let us write
\begin{equation}
Z_{\text{pure}}(\etap)=\frac{1}{1-\etap}+\epsilon
\left(\frac{\etap}{1-\etap}\right)^2,\quad b_3=1+\epsilon,
\label{B.0}
\end{equation}
and set $\epsilon\to 0$ at the end of the calculations. In that
case, replacement into Eq.\ (\ref{new2}) gives
\beq
Z^{\text{SYH}}(\rho)=1+\frac{\eta}{1-\eta}\frac{1}{\langle\sigma\rangle^2}\left[\langle\sigma\rangle
\bar{B}_2+\frac{\eta}{1-\eta}\left(\bar{B}_3-\langle\sigma\rangle
\bar{B}_2\right)\right],
\label{B.1bis}
\eeq
which, for a binary
mixture, becomes
\beq
Z^{\text{SYH}}(\rho)=\frac{1}{1-\eta}\left(1+ x_1 x_2
\frac{\sigma_1+\sigma_2}{\langle\sigma\rangle}\frac{\eta}{1-\eta}\Delta\right).
\label{B.1}
\eeq
Note that Eq.\ (\ref{B.1}) is equivalent to a
series expansion of the exact solution in powers of $\Delta$
truncated in the linear term. In fact, in view of Eqs.\
(\ref{A11.1})--(\protect\ref{A13}), it is exact up to order
$O(\Delta^2)$. Also, it is important to point out that Eqs.\
(\ref{B.1bis}) and (\ref{B.1})  hold regardless of the value of
$\epsilon$, so the limit $\epsilon\to 0$ has not been needed.

As for Hamad's approximation, we would have
\begin{equation}
X_{11}=1+\frac{\sigma_1+\sigma_2}{\langle\sigma\rangle}x_2\Delta,\quad
X_{12}= 1,
\label{B.2}
\end{equation}
 and the similar result for
$X_{22}$ obtained from $X_{11}$ in Eq.\ (\ref{B.2}) with the usual
replacement $1 \leftrightarrow 2$. After some algebra, one finds
\begin{widetext}
\begin{equation}
Z^{\text{H}}=\frac{1}{1-\eta}\left\{1+ x_1 x_2
\frac{\sigma_1+\sigma_2}{\langle\sigma\rangle}\eta\left[1+\eta\left(
\frac{x_1
\sigma_1/\langle\sigma\rangle}{1-\eta\left(1+\frac{\sigma_1+\sigma_2}
{\langle\sigma\rangle}x_2\Delta\right)} +\frac{x_2
\sigma_2/\langle\sigma\rangle}{1-\eta\left(1+\frac{\sigma_1+\sigma_2}
{\langle\sigma\rangle}x_1\Delta\right)}\right)\right]\Delta\right\}.
\label{B.3}
\end{equation}
We remark that Eq.\ (\ref{B.3}) is exact to first order in
$\Delta$.
\end{widetext}

A comparison of the exact compressibility factor with our
approximation (\ref{B.1}) and Hamad's approximation (\ref{B.3})
indicates that Eq.\ (\ref{B.1}), being far simpler than Eq.\
(\ref{B.3}), is better than the latter for $\Delta>0$, both
approaches being comparably good for $\Delta<0$. This is
illustrated in Fig.\ \ref{fig1D}, where we display the exact $Z$
as a function of the non-additivity parameter for a symmetric
($\sigma_2/\sigma_1=1$) and an asymmetric ($\sigma_2/\sigma_1=2$)
binary mixture
 of the same packing fraction
$\eta=0.5$, and mole fraction $x_1=0.25$, together with the two
theoretical approximations.
\begin{figure}[tbp]
\includegraphics[width=0.9\columnwidth]{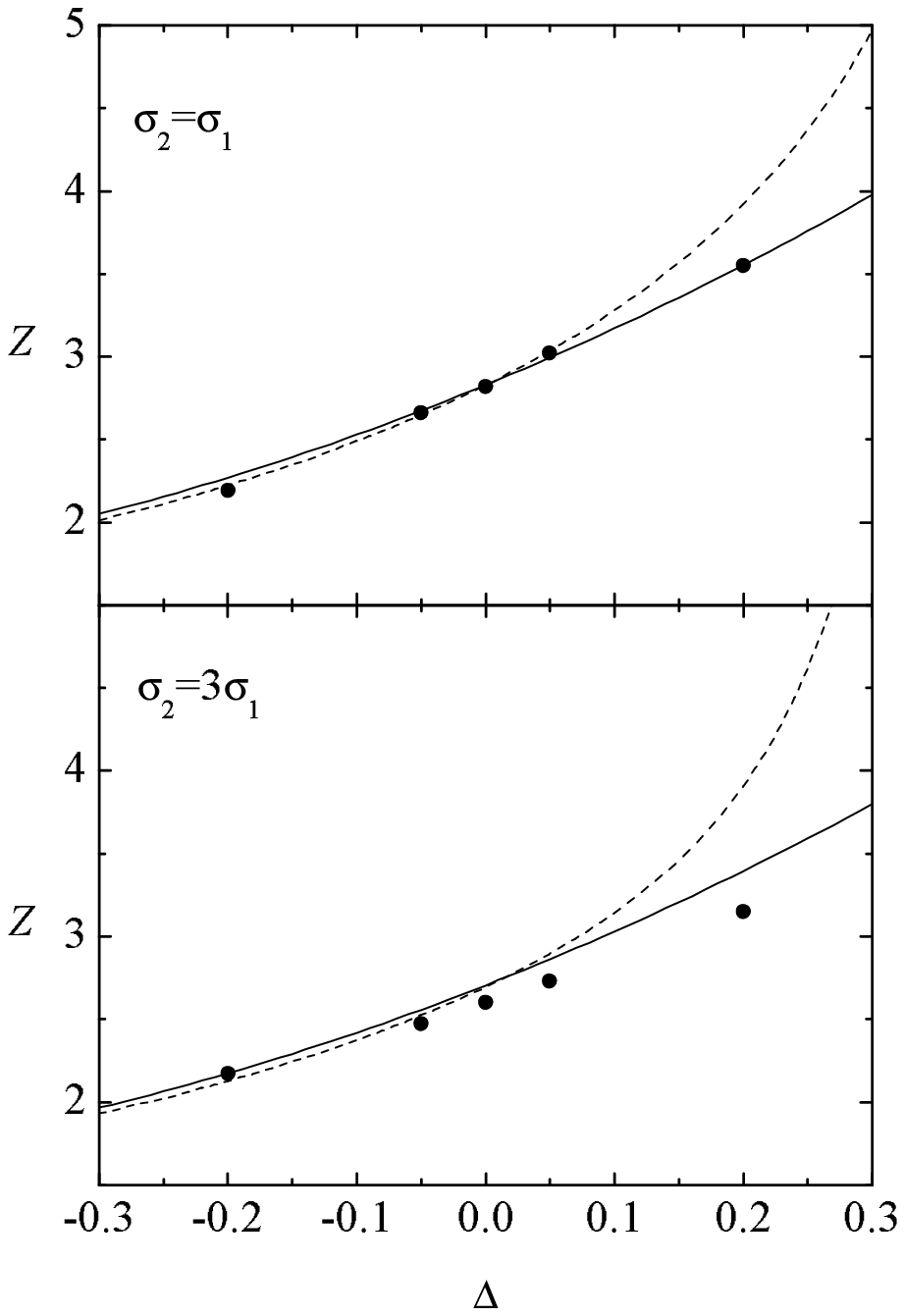}
\caption{Plot of the compressibility factor versus the
non-additivity parameter $\Delta$ for an equimolar symmetric binary
mixture of non-additive hard disks at a packing fraction $\eta=0.4$
(upper panel) and for an equimolar asymmetric mixture with
$\sigma_2/\sigma_1=3$ at $\eta=0.4$ (lower panel). The solid lines
are our proposal, Eq.\ (\protect\ref{new2}), and the dashed lines
are Hamad's proposal, Eq.\ (\protect\ref{H.22}). The circles are
results from molecular dynamics
simulations.\protect\cite{H99}\label{figdisks}}
\end{figure}
\begin{figure}[tbp]
\includegraphics[width=0.9\columnwidth]{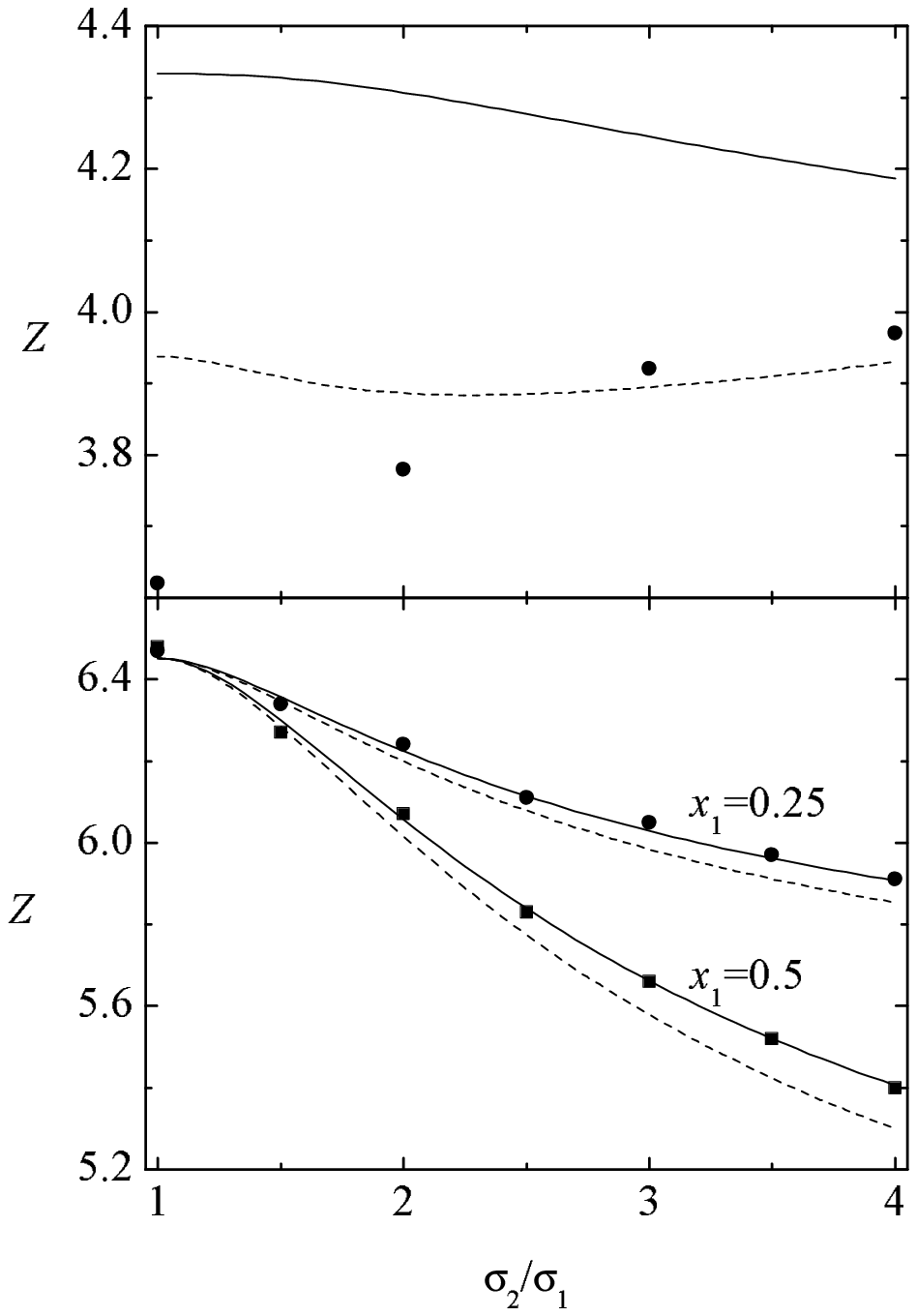}
\caption{Plot of the compressibility factor versus the size ratio
$\sigma_{2}/\sigma_{1}$ for an equimolar binary mixture of
non-additive hard disks with $\Delta=-0.2$ at $\eta=0.6$ (upper
panel) and for two binary additive hard-disk mixtures ($\Delta=0$)
at $\eta=0.6$ (lower panel). The solid lines are our proposal, Eq.\
(\protect\ref{new2}), and the dashed lines are Hamad's proposal,
Eq.\ (\protect\ref{H.22}). The symbols are results from molecular
dynamics simulations.\protect\cite{H99}\label{figdiscosS}}
\end{figure}
\begin{figure}[tbp]
\includegraphics[width=0.9\columnwidth]{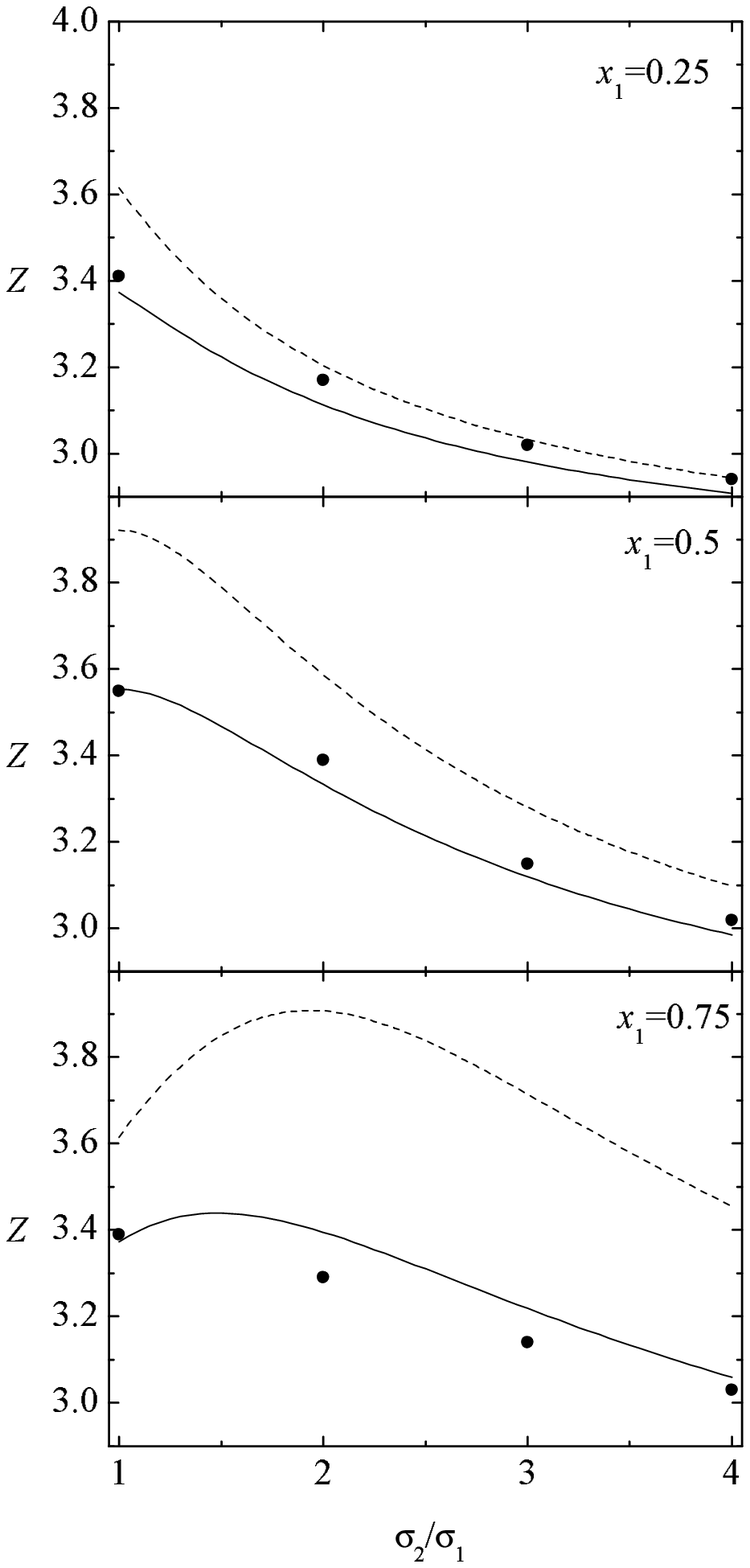}
\caption{Plot of the compressibility factor versus the size ratio
$\sigma_{2}/\sigma_{1}$ for three binary mixtures of non-additive
hard disks with $\Delta=0.2$ at $\eta=0.4$ and $x_1=0.25$ (upper
panel), $x_1=0.5$ (middle panel), and $x_1=0.75$  (lower panel). The
solid lines are our proposal, Eq.\ (\protect\ref{new2}), and the
dashed lines are Hamad's proposal, Eq.\ (\protect\ref{H.22}). The
circles are results from molecular dynamics
simulations.\protect\cite{H99}\label{figdisks3}}
\end{figure}

\subsubsection{Disks ($d=2$) \label{disks}}

It seems natural to begin with the case of symmetric binary
mixtures, i.e., mixtures where $\sigma_1=\sigma_2$, and to
investigate the effect of non-additivity. Representative results in
this respect for an equimolar symmetric binary mixture of
non-additive hard disks are displayed in the upper pannel of Fig.\
\ref{figdisks}, where we have plotted $Z$ as a function of the
non-additivity parameter $\Delta$ at a packing fraction $\eta=0.4$.
 A similar plot of $Z$ versus $\Delta$ is
presented in the lower panel of Fig.\ \ref{figdisks}, but in this
case for an equimolar asymmetric mixture ($\sigma_{2}/\sigma_{1}=3$)
at the same packing fraction $\eta=0.4$.

The size ratio dependence of the compressibility factor is
displayed in Figs.\ \ref{figdiscosS} and \ref{figdisks3} for
various combinations of mole fraction $x_1$, non-additivity
parameter $\Delta$, and packing fraction $\eta$.

Although in the paper by Al-Naafa \textit{et al.}\cite{H99} they
evaluated $Z^{\text{H}}$ by taking for $Z_{\text{pure}}$ the one
that follows from our own simple EOS for the hard-disk
fluid,\cite{SHY95} in Figs.\ \ref{figdisks}--\ref{figdisks3}  we
have considered for both proposals perhaps the most accurate EOS
available nowadays, namely the one due to Luding,\cite{Luding}
\beq
Z_{\text{pure}}^{\text{Luding}}(\etap)= \frac{ 1+ \etap^2/8}{\left(
1-\etap \right)^{2}}-\frac{\etap^4}{ 64 \left( 1-\etap \right)^{4}}.
\label{Luding}
\eeq

Once again we find that the trend observed in $d=1$ is also
present in the case $d=2$, namely, that in general our proposal
performs better than Hamad's,  except for negative $\Delta$. It
is worth recalling here that Hamad's EOS includes the {\em exact}
third virial coefficient, Eqs.\ (\ref{H.9})--(\ref{H.13}), while
ours makes use of the approximation embodied by Eq.\
(\ref{H.13.2}).
\begin{figure}[tbp]
\includegraphics[width=0.9\columnwidth]{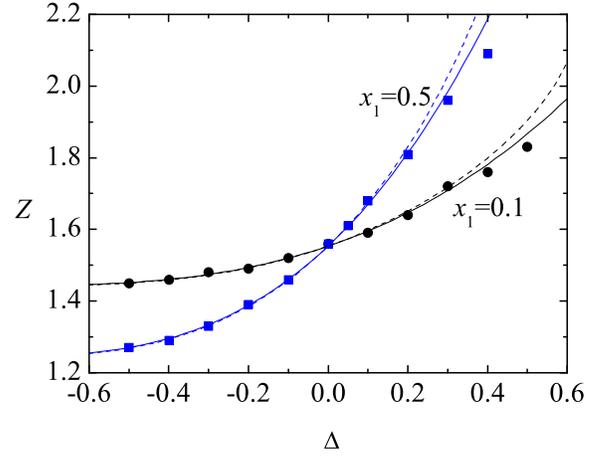}
\caption{Plot of the compressibility factor versus the
non-additivity parameter $\Delta$ for a symmetric binary mixture of
non-additive hard spheres at $\eta=\pi/30$ and two different
compositions. The solid lines are our  proposal, Eq.\
(\protect\ref{new2}), and the dashed lines are Hamad's proposal,
Eq.\ (\protect\ref{H.22}). The symbols are results from Monte Carlo
simulations.\protect\cite{JJR94a,JJR94b}\label{fig3}}
\end{figure}
\begin{figure}[tbp]
\includegraphics[width=0.9\columnwidth]{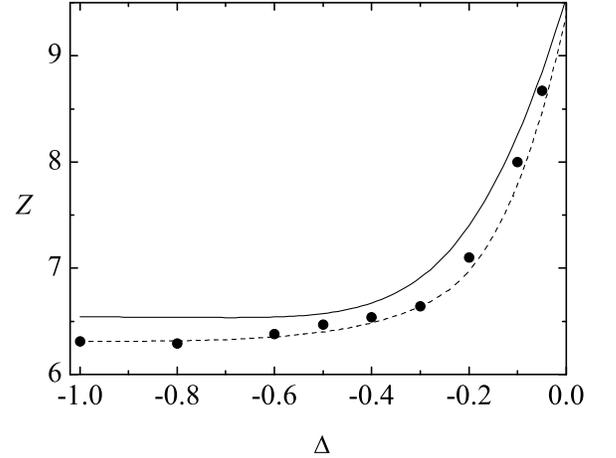}
\caption{Plot of the compressibility factor versus the
non-additivity parameter $\Delta$ for an equimolar asymmetric binary
mixture of non-additive hard spheres with size ratio
$\sigma_{2}/\sigma_{1}=3$, at  $\eta=0.5$. The solid line is our
proposal, Eq.\ (\protect\ref{new2}), and the dashed line is Hamad's
proposal, Eq.\ (\protect\ref{H.22}). The circles are results from
Monte Carlo simulations.\protect\cite{H97}\label{figesferas2Delta}}
\end{figure}
\begin{figure}[tbp]
\includegraphics[width=0.9\columnwidth]{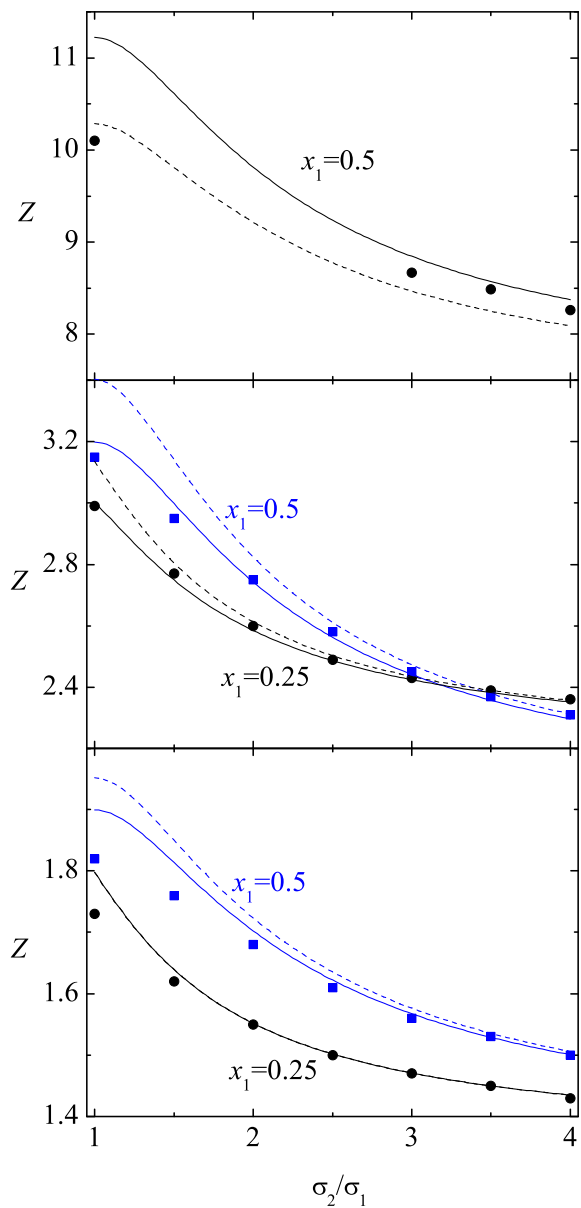}
\caption{Plot of the compressibility factor versus the size ratio
$\sigma_{2}/\sigma_{1}$ for binary mixtures of non-additive hard
spheres with $x_1=0.5$, $\Delta=-0.05$, $\eta=0.5$ (upper panel),
$x_1=0.25, 0.5$, $\Delta=0.2$, $\eta=0.2$ (middle panel), and
$x_1=0.25, 0.5$, $\Delta=0.5$, $\eta=0.075$
 (lower panel). The solid lines are our proposal, Eq.\
(\protect\ref{new2}), and the dashed lines are Hamad's proposal,
Eq.\ (\protect\ref{H.22}). The symbols are results from Monte Carlo
simulations.\protect\cite{H97}\label{figesferasS}}
\end{figure}
\subsubsection{Spheres ($d=3$) \label{spheres}}

We proceed here as in the case of $d=2$. Figure \ref{fig3} shows $Z$
as a function of $\Delta$ for a symmetric binary mixture of
non-additive hard spheres at the packing fraction $\eta=\pi/30\simeq
0.105$ and for $x_1=0.1$ and $x_1=0.5$. Here, as in all the rest of
the calculations for hard-spheres, $Z_{\text{pure}}$ is the one
corresponding to the {Carnahan--Starling--Kolafa (CSK) EOS,
\cite{Ko86}}
\beq
Z_{\text{pure}}^{\text{CSK}}(\etap)=\frac{1+\etap+\etap^2-2\etap^3(1+\etap
)/3}{\left(1-\etap\right)^{3}}.
\label{CS}
\eeq

In Fig.\ \ref{figesferas2Delta} we present a plot of $Z$ versus
$\Delta$, but in this case for an equimolar asymmetric non-additive
hard-sphere mixture  with $\sigma_{2}/\sigma_{1}=3$  at the the
packing fraction $\eta=0.5$. Finally, Fig.\ \ref{figesferasS} is a
plot of $Z$ as a function of the size ratio for different values of
$x_1$, $\Delta$, and density. Once more these figures indicate that
our proposal in the case of $d=3$ is superior to Hamad's, save for
negative non-additivity.

\section{Concluding remarks \label{sec6}}
In this paper we have introduced a new proposal for the EOS of a
multicomponent mixture of $d$-dimensional non-additive hard
spheres. This proposal is an immediate generalization of the one
(rather accurate) we developed for additive hard spheres to which
it immediately reduces if the non-additivity parameters are set
equal to zero. A general prescription for the $d$-dimensional
composition independent third virial coefficients of non-additive
hard-sphere mixtures has also been introduced. It is exact for
$d=1$ and $d=3$ and does a very good job also for $d=2$. In the
absence of exact results or simulation data for other
dimensionalities, its merits in this respect remain to be
evaluated.

Our proposal for the EOS involves providing (sensible)
approximations for the contact values of the radial distribution
functions {that fulfill a few simple requirements. On the one hand,
they reduce to the pure component value $g_{\text{pure}}$ in the
appropriate limit and also comply with the limit in which one of the
species is made of point particles that do not occupy volume. On the
other hand, they yield the exact $g_{ij}$ to first order in the
density.}  Operationally, our proposed EOS for the non-additive
mixture [cf. Eq.\ (\ref{new2})] is given explicit in terms of the
pure component EOS, and the second and third virial coefficients of
the mixture. The former  feature is shared with other proposals in
the literature.\cite{H97,H96b,H96a,H00} In any case, we find that
the present EOS does a good job also in the non-additive situation,
and represents a reasonable compromise between simplicity and
accuracy. {In comparison with Hamad's approach, which is also simple
and reasonably accurate and which we have generalized here to
arbitrary dimensionality, it has the advantage of being able to deal
with asymmetric mixtures where the former faces greater
difficulties.}

Because the full assessment of our proposal involves so many facets,
there are of course many issues that we have not addressed. We have
only attempted to illustrate some of the consequences of employing
our approximate EOS. The results in the previous section illustrate
a trend that we have observed with other values of the parameters,
namely that in general Hamad's proposal does a better job for
negative non-addititivities (especially as the density is increased)
while ours should be preferred in the case of positive
non-additivities, at least for $d=1$, $d=2$, and $d=3$.
Nevertheless, one can see that the performance of our EOS is
reasonably good in highly asymmetric mixtures, even for negative
$\Delta$. {So in some sense, rather than strictly competing, our
approach and Hamad's are complementary.} It is also worth noting
that here we have chosen to take our original recipe of the additive
case\cite{SYH99} for simplicity, but we could have as well
considered the more refined ones that we introduced
later,\cite{SYH02} at the expense of more complicated final
expressions. {Also, the choice of $Z_{\text{pure}}$ is free and the
results of course depend on that choice. Nevertheless, provided
$Z_{\text{pure}}$ is reasonably accurate, the qualitative trends
should not be altered by different choices and this is actually the
case. For instance, in the analysis of non-additive hard disks we
took for $Z_{\text{pure}}$ the one corresponding to Luding's
EOS.\cite{Luding} With minor numerical differences, very similar
results are obtained if Henderson's equation\cite{H75} or our EOS,
\cite{SHY95} which are both accurate, are used instead.}
{Analogously, in the three-dimensional case the results are
practically the same if the Carnahan--Starling EOS\cite{CS69} is
used instead of  Eq.\ (\ref{CS}).}

We are fully aware that interesting features such as the demixing
transition {in the case of positive nonadditivity (both for
symmetric and asymmetric mixtures) remain to be dealt with. We
expect to examine some of these in the future. In any event,
irrespective of the illustrative calculations that we have presented
in this paper, we have attempted to include a rather comprehensive
account of previous work on the subject which will hopefully serve
to provide some perspective and be useful to other researchers}.

\acknowledgments This work has been supported by Junta de
Extremadura  (Consejer\'{\i}a de Educaci\'on y Tecnolog\'{\i}a) and
Fondo Social Europeo under Project TEM04/0009. M.L.H. also wants to
thank D.G.A.P.A.-U.N.A.M. for a sabbatical grant. A.S. and S.B.Y.
acknowledge the financial support of Ministerio de Educaci\'on y
Ciencia (Spain) through Grant No. FIS2004-01399 (partially financed
by FEDER funds).

\appendix
\section{Exact solution in the one-dimensional binary mixture case\label{appA}}
In the one-dimensional case ($d=1$) with nearest-neighbor
interactions [which implies that $2\sigma_{12}\geq
\text{max}\left(\sigma_1, \sigma_2\right)$], the thermodynamic and
structural properties of the binary mixture are exactly
known.\cite{PL54,K55,LZ71,CB99,Penrose} The EOS relating the density
$\rho$ to the pressure $p$ (in units of $k_BT$) and to the diameters
$\sigma_1$, $\sigma_2$, and
$\sigma_{12}=\frac{1}{2}(\sigma_1+\sigma_2)(1+\Delta)$ is given by
\begin{equation}
\frac{1}{\rho}=\frac{1}{p}+\alpha\frac{\sqrt{1+4x_1
x_2(e^{2\alpha p}-1)}-1}{e^{2\alpha p}-1}+\langle\sigma\rangle,
\label{A1new}
\end{equation}
where $\alpha\equiv
\sigma_{12}-(\sigma_1+\sigma_2)/2=(\sigma_1+\sigma_2)\Delta/2\geq
-\text{min}\left(\sigma_1/2, \sigma_2/2\right)$. Note that if
$p\to\infty$ then $\eta\to 1$ for $\alpha>0$, while $\eta\to
[1-2\frac{|\alpha|}{\langle\sigma\rangle} \text{min}(x_1,
x_2)]^{-1}$  for $\alpha<0$.

Equation (\ref{A1new}) can alternatively be written as
\begin{equation}
p=\frac{\rho}{1-\eta}\Phi\left(x_1,\frac{\rho
\alpha}{1-\eta}\right), \label{A10}
\end{equation}
where $\Phi(x_1,\yy)$ is the solution to
\begin{equation}
\Phi^{-1}=1+\yy\frac{1-\sqrt{1+4x_1
x_2(e^{2\yy\Phi}-1)}}{e^{2\yy\Phi }-1} \label{A11a}
\end{equation}
or, equivalently,
\begin{equation}
e^{2\yy\Phi}-1=2\yy\frac{2x_1 x_2 \yy
-(1-\Phi^{-1})}{(1-\Phi^{-1})^2}.
\label{A11b}
\end{equation}

In principle, the compressibility factor is a function of four
parameters: the number density $\rho$, the mole fraction $x_1$, the
size ratio $\sigma_2/\sigma_1$, and the non-additivity parameter
$\alpha$. However, the scaling relation (\ref{A10}) shows that there
are only two independent parameters: the mole fraction $x_1$ and the
scaled parameter $\yy\equiv \rho\alpha/(1-\eta)$. More specifically
[{cf.} Eq.\ (\ref{A10})],
\begin{equation}
Z(\rho)=\frac{1}{1-\eta}\Phi\left(x_1,\frac{\rho \alpha}{1-\eta}
\right).
\label{A11.1}
\end{equation}
Thus, $\Phi(x_1,\yy)$ measures the compressibility factor of the
non-additive mixture, relative to that of an additive mixture with
the same packing fraction.

The expansion of the scaling function $\Phi(x_1,\yy)$ in powers of
$\yy$ is
\begin{equation}
\Phi(x_1,\yy)=\sum_{n=0}^\infty \Phi_n(x_1) \yy^n,
\label{A12}
\end{equation}
where the first few terms are
\begin{equation}
\begin{array}{l}
\Phi_0=1,\quad \Phi_1=2x_1 x_2,\\
\Phi_2=0,\quad \Phi_3=-4 x_1^2x_2^2,\\
\Phi_4=-\frac{8}{3} x_1^2 x_2^2, \quad \Phi_5=-\frac{4}{3} x_1^2
x_2^2(1-8x_1 x_2).
\label{A13}
\end{array}
\end{equation}
In the limit of very small non-additivity, we can make the linear
approximation $\Phi(x_1,\yy)\approx 1+2x_1 x_2 \yy$. This is a good
approximation in the range $-0.4\leq \yy\leq 0.4$. The asymptotic
behaviors of $\Phi(x_1,\yy)$ are easily derived from
Eqs.~(\ref{A11a}) and (\ref{A11b}). In the limit $\yy\to \infty$
(with $\alpha>0$), we simply have $\Phi(x_1,\yy)\to 1$, while in the
limit $\yy\to -\infty$, the result is $\Phi(x_1,\yy)\to -K(x_1)
\yy^{-1}$, where $K(x_1)$ is the solution to
\begin{equation}
4x_1 x_2
K=1+\sqrt{1-4x_1 x_2(1-e^{-2K})}.
\label{A15}
\end{equation}
Note also that $\Phi(x_1,\yy)$ is a non-monotonic function of $\yy$
which presents a maximum for a certain value $\yy_0(x_1)>0$.

From Eqs.~(\ref{A11.1})--(\ref{A13}) it follows that the (exact)
second and third virial coefficients can be written as
\begin{equation}
\bar{B}_2=\langle \sigma\rangle+2x_1x_2\alpha, \quad
\bar{B}_3=\langle \sigma\rangle \left(\langle\sigma\rangle+4
x_1x_2\alpha \right). \label{A9}
\end{equation}
Further, the fugacity $z_1\equiv e^{\mu_1}$ (where $\mu_1$ is the
chemical potential of species $1$, again in units of $k_BT$) is
given by the following expression \beq z_1=\lambda_1 p
e^{\sigma_1 p}\left(1-\frac{\Phi-1}{2x_1 \yy \Phi}\right),
\label{A16} \eeq and a similar expression for $z_2$.

\subsection{Absence of phase separation}
Given the values of $\sigma_1$, $\sigma_2$, and $\sigma_{12}$ (or
$\alpha$), the thermodynamic state of the mixture is characterized
by the pair $(\rho_1,\rho_2)$ or, equivalently, by  $(x_1,
\yy\equiv \rho\alpha/(1-\eta))$. Here we will adopt the latter
viewpoint. If there would exist phase separation into two distinct
phases A and B, the pressure and the chemical potentials should
be equal in both phases. The pressure condition is equivalent to
\begin{equation}
\yy^{\text{A}}\Phi(x_1^{\text{A}},\yy^{\text{A}})=\yy^{\text{B}}\Phi(x_1^{\text{B}},\yy^{\text{B}}).
\label{A17}
\end{equation}
The conditions on the chemical potentials yield
\begin{equation}
\frac{x_1^{\text{A}}}{\Phi(x_1^{\text{A}},\yy^{\text{A}})-1}=
\frac{x_1^{\text{B}}}{\Phi(x_1^{\text{B}},\yy^{\text{B}})-1},
\label{A18a}
\end{equation}
\begin{equation}
\frac{1-x_1^{\text{A}}}{\Phi(x_1^{\text{A}},\yy^{\text{A}})-1}=
\frac{1-x_1^{\text{B}}}{\Phi(x_1^{\text{B}},\yy^{\text{B}})-1}.
\label{A18b}
\end{equation}
These two equations imply
\begin{equation}
x_1^{\text{A}}=x_1^{\text{B}},
\label{A19}
\end{equation}
\begin{equation}
 \Phi(x_1^{\text{A}},\yy^{\text{A}})
=\Phi(x_1^{\text{A}},\yy^{\text{B}}).
\label{A20}
\end{equation}
Given the non-monotonic behavior of $\Phi$ as a function of $\yy$,
Eq.~(\ref{A20}) has solutions with $\yy^{\text{A}}\neq
\yy^{\text{B}}$. However, the combination of (\ref{A17}) and
(\ref{A20}) means that
\begin{equation}
\yy^{\text{A}}=\yy^{\text{B}},
\label{A21}
\end{equation}
and so the only solution is the trivial one.

\subsection{Distribution functions at contact}
From Lebowitz and Zomick's paper\cite{LZ71} (and after some
algebra), one can get the contact values
\begin{equation}
g_{11}=\frac{1}{1-\eta}\frac{2x_1\yy\Phi-(\Phi-1)}{2x_1^2 \yy},
\label{A24a}
\end{equation}
\begin{equation}
g_{22}=\frac{1}{1-\eta}\frac{2x_2\yy\Phi-(\Phi-1)}{2x_2^2 \yy},
\label{A24b}
\end{equation}
\begin{equation}
g_{12}=\sqrt{g_{11}g_{22}}e^{-\yy \Phi}.
\label{A24c}
\end{equation}
Using the expansion (\ref{A12}), one has
 \begin{equation}
g_{11}=\frac{1}{1-\eta}\left[1+2x_2 \yy+2x_2^2 \yy^2
+{O}(\yy^3)\right],
\label{A25a}
\end{equation}
\begin{equation}
g_{22}=\frac{1}{1-\eta}\left[1+2x_1 \yy+2x_1^2 \yy^2
+{O}(\yy^3)\right],
\label{A25b}
\end{equation}
\begin{equation}
g_{12}=\frac{1}{1-\eta}\left[1-2x_1x_2 \yy^2 +{O}(\yy^3)\right].
\label{A25c}
\end{equation}

\section{Some special limits \label{appB}}
It is  interesting to examine the performance of Hamad's
approximation,
 Eqs.\ (\ref{H.19}) and (\ref{H.22}), and of our proposal,
Eqs.\ (\ref{C1bis}) and (\ref{new2}), in the following special
limits.

\subsection{$\sigma_{12}=0$}
In the limit of extreme negative non-additivity ($\sigma_{12}= 0$ or
$\Delta=-1$), one has $\widehat{\sigma}_{1}=\widehat{\sigma}_{2}=0$,
so that Eq.\ (\ref{negative}) yields
\begin{equation}
\begin{array}{l}
c_{1;11}=\frac{b_3}{b_2}\sigma_1^d,\quad
c_{2;22}=\frac{b_3}{b_2}\sigma_2^d,\\
c_{1;12}=c_{1;22}=c_{2;11}=c_{2;12}=0.
\end{array}
\label{new3}
\end{equation}
The above expressions are exact in that limit. Hamad's proposal
(\ref{H.19}) becomes then
\begin{equation}
X_{11}=\frac{x_1\sigma_1^d}{\langle\sigma^d\rangle},\quad
X_{22}=\frac{x_2\sigma_2^d}{\langle\sigma^d\rangle},\quad X_{12}=0,
\label{new4}
\end{equation}
\begin{equation}
g_{11}^{\text{H}}(\rho) =g_{\text{pure}}\left(\eta_1\right), \quad
g_{22}^{\text{H}}(\rho) =g_{\text{pure}}\left(\eta_2\right), \quad
g_{12}^{\text{H}}(\rho)=1, \label{23b}
\end{equation}
where $\eta_i=v_d\rho x_i\sigma_i^d$ is the partial packing fraction
of species $i$. Equation (\ref{23b}) is the exact result, reflecting
the fact that in the limit $\sigma_{12}= 0$ the mixture is actually
made of two mutually independent one-component fluids. On the other
hand, in our proposal we have
\begin{equation}
z_{11}=\frac{b_3x_1\sigma_1^d-b_2\langle\sigma^d\rangle}{(b_3-b_2)\langle\sigma^d\rangle},\quad
z_{12}=-\frac{b_2}{b_3-b_2},
\label{new5}
\end{equation}
\begin{equation}
g_{11}^{\text{SYH}}(\rho)=\frac{1}{1-\eta}\frac{b_3x_2\sigma_2^d}{(b_3-b_2)
\langle\sigma^d\rangle}+g_{\text{pure}}(\eta)\frac{b_3x_1\sigma_1^d-b_2
\langle\sigma^d\rangle}{(b_3-b_2)\langle\sigma^d\rangle},
\label{new6}
\end{equation}
\begin{equation}
g_{12}^{\text{SYH}}(\rho)=\frac{1}{1-\eta}\frac{b_3}{b_3-b_2}-
g_{\text{pure}}(\eta)\frac{b_2}{b_3-b_2},
 \label{new6bis}
\end{equation}
plus the equivalent expressions obtained by the exchange
$1\leftrightarrow 2$. Equations(\ref{new6}) and (\ref{new6bis})
are only exact to first order in the density.

\subsection{Widom--Rowlinson limit}
The WR limit ($\sigma_1=\sigma_2\to 0$) represents an extreme case
of positive non-additivity ($\Delta\to\infty$). The coefficients
$c_{k;ij}$ are
\begin{equation}
c_{1;11}=c_{1;12}=c_{2;12}=c_{2;22}=0,\quad
c_{1;22}=c_{2;11}=(2\sigma_{12})^d. \label{new7}
\end{equation}
In this WR limit the packing fraction vanishes, so that the relevant
density parameter is $\eta'=v_d\rho\sigma_{12}^d$. In Hamad's
approximation,
\begin{equation}
\langle\sigma^d\rangle
X_{11}=\frac{b_2^2}{2b_3}x_2\sigma_{12}^d,\quad
\langle\sigma^d\rangle X_{12}=0,
\label{new8}
\end{equation}
\begin{equation}
g_{11}^{\text{H}}(\rho)=g_{\text{pure}}\left(\frac{b_2^2}{2b_3}x_2\eta'\right),\quad
g_{12}^{\text{H}}(\rho)=1. \label{new9}
\end{equation}
Our approximation yields
\begin{equation}
\langle\sigma^d\rangle
z_{11}=\frac{b_2^2x_2\sigma_{12}^d}{2(b_3-b_2)},\quad
\langle\sigma^d\rangle z_{12}=0,
\label{new9a}
\end{equation}
\begin{equation}
g_{11}^{\text{SYH}}(\rho)=1+\frac{b_2}{2}x_2\eta',\quad
g_{12}^{\text{SYH}}(\rho)=1. \label{new10}
\end{equation}
Both approximations differ in $g_{11}(\rho)$ and $g_{22}(\rho)$, but
these contact values do not contribute to $Z(\rho)$ in the WR limit.
The result is in the two cases
\begin{equation}
Z(\rho)=1+2^d x_1 x_2 \eta', \label{17}
\end{equation}
which is the mean field result.

We note that in the one-dimensional case the exact result that
follows when setting $\sigma_1=\sigma_2=0$, $\eta'=\rho \alpha$ in
Eq.\ (\protect\ref{A11.1}) is
\begin{equation}
Z(\rho)=\Phi\left(x_1,{\eta'}\right)\quad (d=1). \label{A11.3}
\end{equation}

\subsection{Asakura--Oosawa limit}
The Asakura--Oosawa limit consists of setting $\sigma_2=0$ and
$\sigma_{12}=\sigma_1/2+R$, where $R$ represents the radius of
gyration. In that case, $\widehat{\sigma}_1=\sigma_1+2R$ and
$\widehat{\sigma}_2=2R$, so that Eq.\ (\ref{negative}) gives
\beq
\begin{array}{l}
c_{1;11}=\frac{b_3}{b_2}\sigma_1^d,\quad
c_{2;11}=(2R)^d+\left(\frac{b_3}{b_2}-1\right)\sigma_1(2R)^{d-1},\\
 c_{1;22}=(\sigma_1+2R)^d,\quad c_{2;22}=0,\\
c_{1;12}=\sigma_1^d+\left(\frac{b_3}{b_2}-1\right)\frac{4R\sigma_1^d}{\sigma_1+2R},\quad
c_{2;12}=0.
 \end{array}
 \label{new13}
\end{equation}
From (\ref{new13}), it follows that
\begin{equation}
z_{11}=1+\frac{x_2}{x_1}\frac{b_2}{b_3-b_2}\left(\frac{2R}{\sigma_1}\right)^{d-1}
\left(\frac{2R}{\sigma_1}+\frac{b_3}{b_2}-1\right),
\label{new14}
\end{equation}
\begin{equation}
z_{22}=\frac{b_2}{b_3-b_2}\left[\left(1+\frac{2R}{\sigma_1}\right)^d-1\right],
\label{new15}
\end{equation}
\begin{equation} z_{12}=\frac{4R/\sigma_1}{1+2R/\sigma_1}.
\label{new16}
\end{equation}
Further, in this limit $\langle\sigma^d\rangle=x_1 \sigma_1^d$,
$\bar{B}_2= 2^{d-1} x_1 \left[ x_1 \sigma_1^d  +2\left( 1-x_1
\right) \left({\sigma_1}/{2} + R \right)^d \right]$ and $\bar{B}_3$
may be computed from the $c_{k;ij}$ given in (\ref{new13}).
Therefore, upon substitution into Eq.\ (\ref{new2}), one would get
the EOS for the Asakura--Oosawa limit. Since the resulting
expression is not very illuminating, it will be omitted. Similarly,
with the substitution of Eqs.\ (\ref{new14})--(\ref{new16}) into
Eq.\ (\ref{C1bis}) the contact values of the radial distribution
functions (which will be also omitted) follow. The corresponding
results for this limit in Hamad's proposal are readily derived from
Eqs.\ (\ref{zX}) and (\ref{new14})--(\ref{new16}),  and subsequent
substitution into Eqs.\ (\ref{H.19}) and (\ref{H.22}).

In $d=1$, taking the Asakura--Oosawa limit ($\sigma_2=0$,
$\alpha=R$) in Eq.~(\protect\ref{A11.1}) we have the exact result
\begin{equation}
Z(\rho)=\frac{1}{1-\eta}\Phi\left(x_1,\frac{\rho R}{1-\eta} \right)
\quad (d=1).
\label{A11.2}
\end{equation}


\end{document}